\documentclass{emulateapj}

\begin{document}
\title{BOSS Ultracool Dwarfs I: Colors and Magnetic Activity of M and L dwarfs}
\shorttitle{BOSS Ultracool Dwarfs I}

\author{Sarah J. Schmidt\altaffilmark{1}, Suzanne L. Hawley\altaffilmark{2}, Andrew A. West\altaffilmark{3}, John J. Bochanski\altaffilmark{4,5}, James R. A. Davenport\altaffilmark{2}, Jian Ge \altaffilmark{6}, and Donald P. Schneider\altaffilmark{7,8}}

\altaffiltext{1} {Department of Astronomy, Ohio State University, 140 West 18th Avenue, Columbus, OH 43210}
\email{schmidt@astronomy.ohio-state.edu}
\altaffiltext{2} {Department of Astronomy, University of Washington, Box 351580, Seattle, WA 98195}
\altaffiltext{3} {Department of Astronomy, Boston University, CAS 422A, 725 Commonwealth Ave, Boston, MA 02215}
\altaffiltext{4} {Rider University, 2083 Lawrenceville Rd, Lawrenceville, NJ 08648, USA}
\altaffiltext{5} {Department of Astronomy, Haverford College, 370 Lancaster Avenue, Haverford, PA 19041}
\altaffiltext{6} {Astronomy Department, University of Florida, Gainesville, FL 32611}
\altaffiltext{7} {Department of Astronomy and Astrophysics, The Pennsylvania State University, University Park, PA 16802}
\altaffiltext{8} {Institute for Gravitation and the Cosmos, The Pennsylvania State University, University Park, PA 16802}

\begin{abstract} 
We present the colors and activity of ultracool (M7-L8) dwarfs from the Tenth Data Release of the Sloan Digital Sky Survey (SDSS). We combine previous samples of SDSS M and L dwarfs with new data obtained from the Baryon Oscillation Sky Survey (BOSS) to produce the BOSS Ultracool Dwarf (BUD) sample of 11820 M7-L8 dwarfs.  By combining SDSS data with photometry from the Two Micron All Sky Survey and the Wide-Field Infrared Sky Explorer mission, we present ultracool dwarf colors from $i-z$ to $W2-W3$ as a function of spectral type, and extend the SDSS-2MASS-WISE color locus to include ultracool dwarfs. The $i-z$, $i-J$, and $z-J$ colors provide the best indication of spectral type for M7-L3 dwarfs. We also examine ultracool dwarf chromospheric activity through the presence and strength of H$\alpha$ emission. The fraction of active dwarfs rises through the M spectral sequence until it reaches $\sim$90\% at spectral type L0. The fraction of active dwarfs then declines to 50\% at spectral type L5; no H$\alpha$ emission is observed in the late-L dwarfs in the BUD sample. The fraction of active L0-L5 dwarfs is much higher than previously observed. The strength of activity declines with spectral type from M7 through L3, after which the data do not show a clear trend. Using one-dimensional chromosphere models, we explore the range of filling factors and chromospheric temperature structures that are consistent with H$\alpha$ observations of M0-L7 dwarfs. M dwarf chromospheres have a similar, smoothly varying range of temperature and surface coverage while L dwarf chromospheres are cooler and have smaller filling factors. 
\end{abstract}


\section{Introduction}
\label{sec:intro}
Ultracool (late-M and L) dwarfs include both the bottom of the hydrogen burning main sequence and the warmest brown dwarfs \citep{Chabrier2000b,Burrows2001}. Spectral types M7--L8 are subject to a variety of changes with decreasing effective temperature: dust clouds make an increasingly important contribution to the atmospheric chemistry \citep{Tsuji1996b,Helling2008b}, diagnostics of magnetic activity indicate changes in the interactions between the magnetic field and ultracool atmosphere \citep{Mohanty2002,Hallinan2008} and the contribution of stars to each spectral type bin decreases as brown dwarfs become the dominant ultracool population. Many of these processes have been investigated with small samples of peculiar or nearby objects, but data from the Sloan Digital Sky Survey \citep[SDSS;][]{York2000} have provided a windfall of information that can be used to understand the bulk properties of ultracool dwarfs, including their colors and magnetic activity. 

A well-defined color locus for ultracool dwarfs is essential to the selection and classification of these objects. Color-spectral type relations can also be used to provide an initial spectral type/effective temperature estimate, prioritizing targets for spectroscopic observations \citep[e.g.,][]{Zhang2009,Castro2012}. Deviations from established color relations are often indicators of gravity \citep{Schmidt2010} or metallicity \citep{Bochanski2013}. While the combination of colors from SDSS and the Two Micron All-Sky Survey \citep[2MASS;][]{Skrutskie2006} provides a broad color space to examine late-M and L dwarfs, the recent data release from the Wide-Field Infrared Sky Explorer \citep[WISE;][]{Wright2010} expands the available color space in which to study ultracool dwarfs. Initial efforts to examine WISE colors have focused on earlier-type main sequence stars \citep{Davenport2014,Theissen2014} or cooler brown dwarfs \citep{Kirkpatrick2011}, so the BUD sample fills an important gap in WISE color sequences.

The spectroscopic component of SDSS also provides a unique opportunity to examine the magnetic activity of a large sample of ultracool dwarfs. Chromospheric activity, which is ubiquitous in mid- to late-M dwarfs, is often classified based on the presence and strength of H$\alpha$ emission \citep[e.g.,][]{Hawley1996,Liebert2003,West2004}. The strength of H$\alpha$, frequently parameterized as the ratio of the luminosity in the H$\alpha$ line to the bolometric luminosity \citep[log($L_{\rm H\alpha}/L_{\rm bol}$); e.g.,][]{Hawley1996}, has an average value that is relatively constant for early-M dwarfs, albeit with a large dispersion, then shows a steady decline through early-L spectral types \citep{Gizis2000,Schmidt2007,Reiners2008,West2008}. 

The fraction of cool and ultracool dwarfs showing detectable H$\alpha$ emission increases from M0 through late-M spectral types and appears to decline at later types \citep{Kirkpatrick1999,Kirkpatrick2000,Gizis2000,West2004}. However, the decline at late-M spectral types has recently been attributed to the difficulty of detecting relatively weak H$\alpha$ emission in these faint objects \citep[e.g.,][]{Schmidt2007,Reiners2008,West2008}, indicating that the fraction of active dwarfs could increase through early-L spectral types. 

Strong radio emission from late-M and early-L dwarfs indicates that ultracool dwarfs are still capable of generating and sustaining strong magnetic fields \citep{Stelzer2006,Berger2010}. A surface magnetic field is a necessary but not sufficient condition for chromospheric activity; the chromosphere must also be heated by the interaction of charged particles with the magnetic field. The decline in the \textit{strength} of H$\alpha$ emission with later spectral type may be the result of increasingly cool photospheres having high magnetic resistivity due to low ionization fractions \citep[e.g.,][]{Mohanty2002}. We use one-dimensional chromosphere models combined with H$\alpha$ data for thousands of M and L dwarfs to investigate the physical characteristics of the average ultracool dwarf chromosphere. Understanding magnetic activity on these objects is essential for selecting targets in next generation doppler surveys for planetary companions \citep{Reiners2010a,Quirrenbach2012}.

The SDSS spectroscopic database, as of the Seventh Data Release \citep[DR7;][]{Abazajian2009}, contained 70,481 M dwarfs \citep[][hereafter W11]{West2011} and 484 L dwarfs \citep[][hereafter S10]{Schmidt2010}. As part of SDSS-III \citep{Eisenstein2011}, the Baryon Oscillation Sky Survey \citep[BOSS;][]{Dawson2013} continued to use the SDSS 2.5-m telescope \citep{Gunn2006} with a similar fiber-fed spectrograph \citep{Smee2013} for additional optical spectroscopy. We were awarded a BOSS ancillary program to target candidate ultracool dwarfs and increase the number of M7 and later dwarfs with optical spectra from SDSS. Over the course of the survey, BOSS obtained spectra of $\sim$10,000 ultracool dwarf candidates. In this paper, we introduce the initial BOSS Ultracool Dwarfs (BUD) sample, which includes data from the first two years of BOSS, corresponding to the Tenth Data Release \citep[DR10;][]{Ahn2014} in addition to late-M (W11) and L (S10) dwarfs from the SDSS DR7. 

The initial target selection and analysis of the present BUD sample is outlined in Section~\ref{sec:sample_BUD}. In Section~\ref{sec:phot}, we describe the  SDSS, 2MASS, and WISE photometry, and in Section~\ref{sec:colors} we examine the ultracool dwarf color locus and the correlation of colors and spectral type. We discuss the presence and strength of H$\alpha$ emission in Section~\ref{sec:haobs} using the BUD data supplemented by previously reported H$\alpha$ detections and upper limits. Section~\ref{sec:tstr} outlines our implementation of the NLTE radiative transfer code, RH \citep{Uitenbroek2001}, that was used to produce a grid of model atmospheres for ultracool dwarfs. In Section~\ref{sec:chr_results}, we compare the ranges of chromospheric temperature structures and filling factors to the data and discuss the results. Finally, in Section~\ref{sec:sum} we provide a summary and an outline of future BUD papers. 

\section{Sample Selection and Spectra}
\label{sec:sample_BUD}
The BUD sample combines data from three different components of SDSS: M7-M9 dwarfs from the DR7 M dwarf sample described by W11, L dwarfs from the DR7 L dwarf sample discussed by S10, and late-M and L dwarfs selected from the BOSS component of DR10. This section describes the selection of these three components.

\subsection{Late-M Dwarfs From DR7}
\label{sec:drm}
The largest component of our sample is a subset of the W11 M dwarf sample selected from DR7. The initial selection criteria used both color ($r-i > 0.42$ and $i-z > 0.24$) and signal-to-noise ratio (S/N $>$3 per 1.9\AA~pixel at $\sim$8300~\AA). Over 100,000 spectra met the color and S/N criteria and were visually assigned spectral types, resulting in 70,841 M dwarfs. From the W11 sample, we select M7-M9 dwarfs that were not flagged as binary systems containing both a white dwarf and an M dwarf (WD-dM pairs), resulting in 9614 late-M dwarfs. We adopt spectral types directly from W11, but re-measure H$\alpha$ equivalent widths (EW) as described in Section~\ref{sec:habud}. We do not use the photometry from the W11 catalog. Instead, we re-query the SDSS and 2MASS catalogs to ensure the entire BUD sample photometry contained uniform flag and uncertainty cuts (described in Section~\ref{sec:phot}). 

\subsection{L Dwarfs From DR7}
\label{sec:drl}
Our sample also includes the 484 L dwarfs selected by S10 from SDSS DR7, using a single color cut of $i-z > 1.4$ and requiring sufficient S/N to assign a type by eye (no specific S/N cut was made; spectra that were too noisy to match to a spectral template were rejected). Again, we adopt the spectral types from S10 and remeasure the H$\alpha$ EWs. We also measure a S/N at $\sim$8300\AA~to identify a subsample of 128 dwarfs that satisfy the S/N$>$3 per pixel criterion for the DR7 M dwarfs. Again, we do not adopt the photometry directly from the S10 sample; instead we re-query the DR10 catalog for updated photometry (see Section~\ref{sec:phot}).

\subsection{Late-M and L dwarfs from BOSS}
While some M and L dwarfs from DR7 were specifically targeted, most of the M and L dwarfs from DR7 were observed primarily because their colors were similar to those of red galaxies and distant quasars. During BOSS observations, we supplemented those serendipitous targets with an ancillary program specifically designed to identify ultracool dwarfs. As described in Section~\ref{sec:selboss}, we selected the DR10 component of the final BUD sample from BOSS data without any reference to their targeting (i.e., point sources targeted by the main survey were treated the same as those targeted by our BUD program). The BUD ancillary program did affect which point sources were targeted (and so ultimately affected our sample) and is described in more detail below. 

\subsubsection{Color Selection of the BOSS Ancillary Targets}
\label{sec:samboss}
Our BOSS ancillary program contains 10,000 spectra, assigned with different surface densities in two regions of the sky. In Stripe82, a 220 deg$^2$ region of the sky around the south Galactic pole \citep{Stoughton2002}, our average targeting density was $\sim$6 deg$^{-2}$. In the remainder of the legacy footprint ($\sim$7430 deg$^2$ towards the northern Galactic pole), our targeting density was $\sim$1 deg$^{-2}$. The primary goal of our target selection was to obtain a spectrum for every L dwarf candidate in the SDSS footprint. 

We began the target selection with a list of all SDSS point sources with good quality $i$ and $z$ photometry and $i-z>1$. In the S10 sample, 97.5\% of the L dwarfs had matches with good quality 2MASS photometry, so the next step in our target selection was to cross-match the list of objects with  2MASS sources using a matching radius of 5\arcsec. Additional color cuts in $i-J$ and $z-J$, listed in Table~\ref{tab:photcut}, excluded objects that fell far from the late-M and L dwarf color locus \citep{Schmidt2010,West2011}. The remaining cuts in $i-z$ and $i$ magnitudes were selected based on the targeting density. The limiting magnitude in the Stripe82 region ($i<21$) was selected based on the estimate of S/N$\sim$5 per pixel (assuming the $\sim$68 km s$^{-1}$ pixels of SDSS spectra) for $i=21$ \citep{Eisenstein2011}. Because we were awarded a lower density of targets in the main (legacy) region, we used a slightly brighter limiting magnitude ($i<20.5$; resulting in a smaller number of possible targets). These magnitude limits should include early-L (L0-L2) dwarfs out to a distance of $\sim$100pc, and late-L dwarfs (L5-L8) to $\sim$15pc (S10).

The $i-z$ color cut for the main SDSS footprint ($i-z>1.44$) was selected to include all L dwarfs; it is located four standard deviations bluer than the L0 dwarf median color at $i-z=1.85$ (S10). A bluer limit ($i-z>1.14$) was placed on the Stripe82 targets, in part to test the color criteria for the main survey, and in part to include a significant number of M8 and M9 dwarfs. At the completion of DR10, just over half (5007) of the total 10,000 targets were observed; 984 of those were repeat observations of late-M and L dwarfs from DR7 to obtain spectra with increased S/N and wavelength coverage. A summary of the selection criteria for the BOSS component of the sample is given in Table~\ref{tab:photcut}.

\begin{deluxetable}{lll} \tablewidth{0pt} \tabletypesize{\scriptsize}
\tablecaption{BOSS Ancillary Target Selection \label{tab:photcut} }
\tablehead{ \colhead{Color or}  & \colhead{Stripe82} &  \colhead{main sample}\\ 
 \colhead{Magnitude}  & \colhead{(SGP)} &  \colhead{(NGP)}} 
\startdata
$i-z$ & $>1.14$ &   $>1.44$ \\
$i$ & $<21$ &   $<20.5$ \\
$i-J$ & $>3.7$ &   $>3.7$ \\
$z-J$ & $1.9<z-J<4$ &    $1.9<z-J<4$ 
\enddata
\end{deluxetable}

\subsubsection{Selecting Ultracool Dwarfs From BOSS Spectra}
\label{sec:selboss}
The ultracool dwarfs selected from DR10 were intended to expand the already existing DR7 samples of late-M and L dwarfs (discussed in Sections~\ref{sec:drm} and~\ref{sec:drl}). We combined the color criteria to select both late-M and L dwarfs. Candidate DR10 ultracool dwarfs either met the criteria for M dwarfs ($r-i > 0.42$ and $i-z > 0.24$) or L dwarfs (which may not have good $r$-band photometry; $i-z>1.4$). These color criteria resulted in a total of 27,967 candidates in all DR10 BOSS plates, each possessing a fully reduced a calibrated spectrum processed using the BOSS pipeline \citep{Bolton2012}.

We visually spectral typed each object using the Hammer spectral typing software \citep{West2004,Covey2007}. Of the initial list, 23,377 were identified as M or L dwarfs (the rejected 4,590 objects include spectra too noisy to assign spectral types, A-K stars, and extragalactic sources). The sample of M7 and later dwarfs in DR10 comprises 3089 spectra. No objects were excluded when a color cut was applied to find WD-dM binaries \citep{Smolcic2004}. For the M7-M9 dwarfs, we included only spectra with S/N$>$3 per pixel at $\sim$8000~\AA, excluding 392 dwarfs for a total of 2516 M7-M9 dwarfs. We included all 181 L dwarfs, 47 of which were flagged as having S/N$<3$. In our final sample, 2239 (89\%) late-M and 111 (61\%) L dwarfs were originally targeted as part of our ancillary program. 

\begin{figure}
\includegraphics[width=\linewidth]{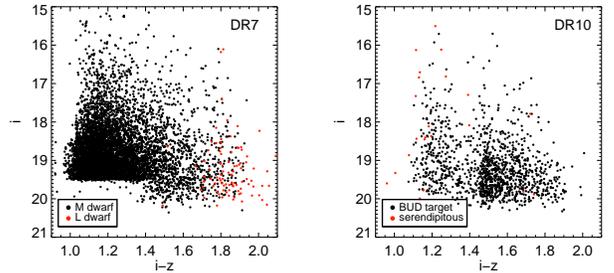} 
\caption[Hess diagrams of the combined DR7 M7-M9 and L dwarf samples compared to the BUD sample.]{Hess diagrams (apparent $i$ magnitude as a function of $i-z$ color) for the DR7 M7-M9 and L dwarf samples (from W11 and S10; left) and the component of the BUD sample selected from DR10 (right). In the DR7 sample, we differentiate between M and L dwarfs, while in the DR10 sample, we distinguish dwarfs targeted specifically for the sample (black dots; as described in Section~\ref{sec:samboss}) from those observed due to other target selections (red dots). The blue limit of the DR7 sample reflects the spectral type cut at M7, while the excluded color and magnitude regions represent the criteria applied to select low and high redshift quasars \citep{Richards2002}. The artifacts at $i-z$ = 1.14 and $i-z$ = 1.44 in the DR10 sample (right) are due to the color cuts applied to the Stripe82 and the main sample, respectively (see Section~\ref{sec:samboss} and Table~\ref{tab:photcut}).} \label{fig:hess}
\end{figure}

Hess diagrams of the DR7 and DR10 components of the BUD sample are presented in Figure~\ref{fig:hess}. Each of the diagrams shows artifacts from the color and magnitude cuts applied to select targets for spectroscopic observation. The DR7 late-M dwarfs ($1.0 < i-z < 1.44$) show a strong magnitude limit at $i<19$, while redder objects ($1.44 < i-z < 2.0$) were limited to $z<19$. Those limits were the result of criteria applied to select low and high redshift quasars \citep{Richards2002} rather than cuts for ultracool dwarfs. The color distribution of the DR10 sample is strongly affected by the cuts described in Section~\ref{sec:samboss}. DR10 includes far fewer late-M dwarfs than DR7; those at $i-z<1.44$ are primarily from our Stripe82 targets. For the objects targeted as L dwarfs, the color ($i-z>1.44$) and magnitude ($i<20.5$) limits can be seen clearly. Of the 2967 dwarfs included in the BUD sample, 347 were targeted as part of the main BOSS survey \citep[selection detailed in][]{Dawson2013}.

\section{BUD Data from SDSS, 2MASS, and WISE}
\label{sec:phot}
\begin{figure}
\includegraphics[width=\linewidth]{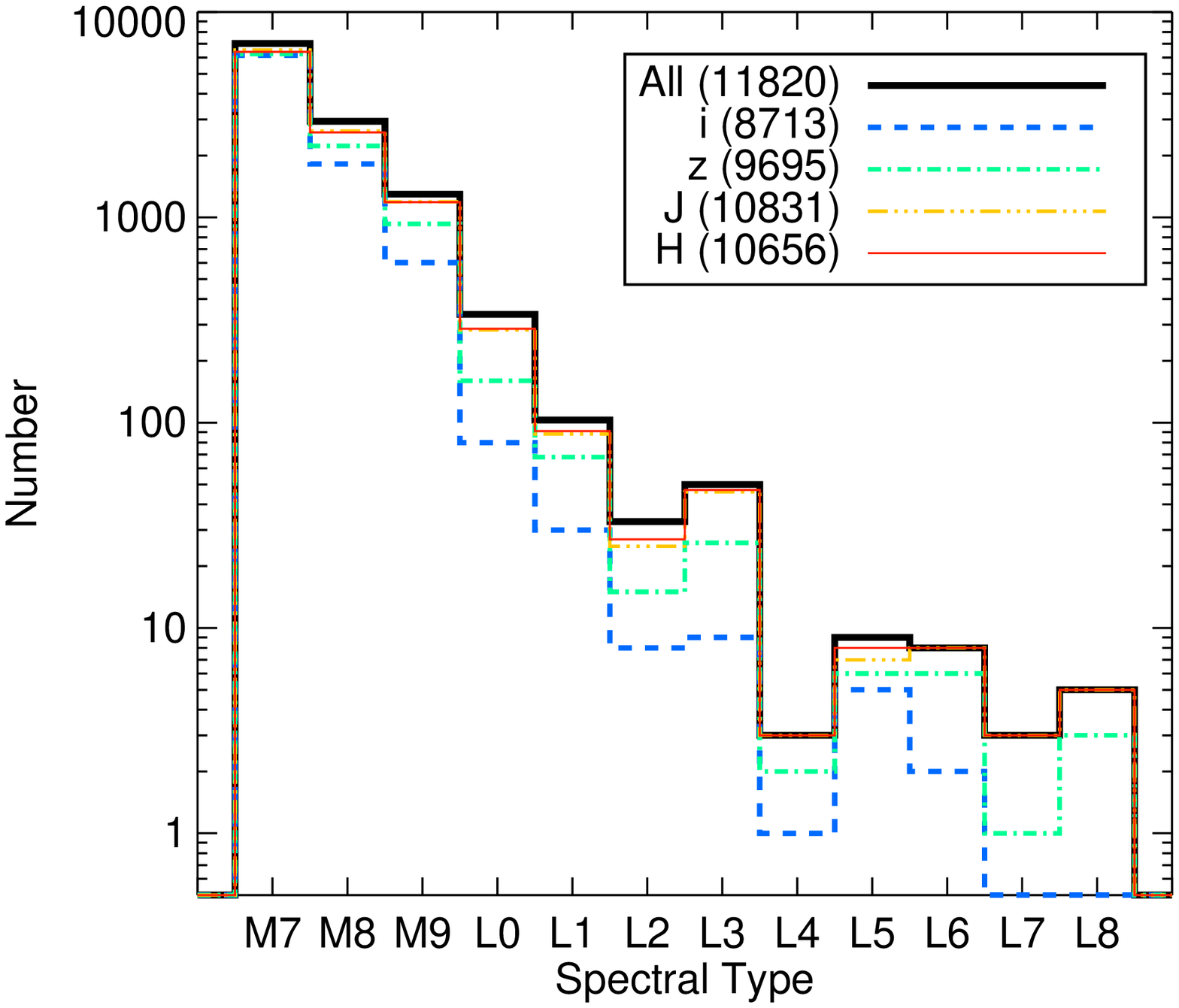} 
\includegraphics[width=\linewidth]{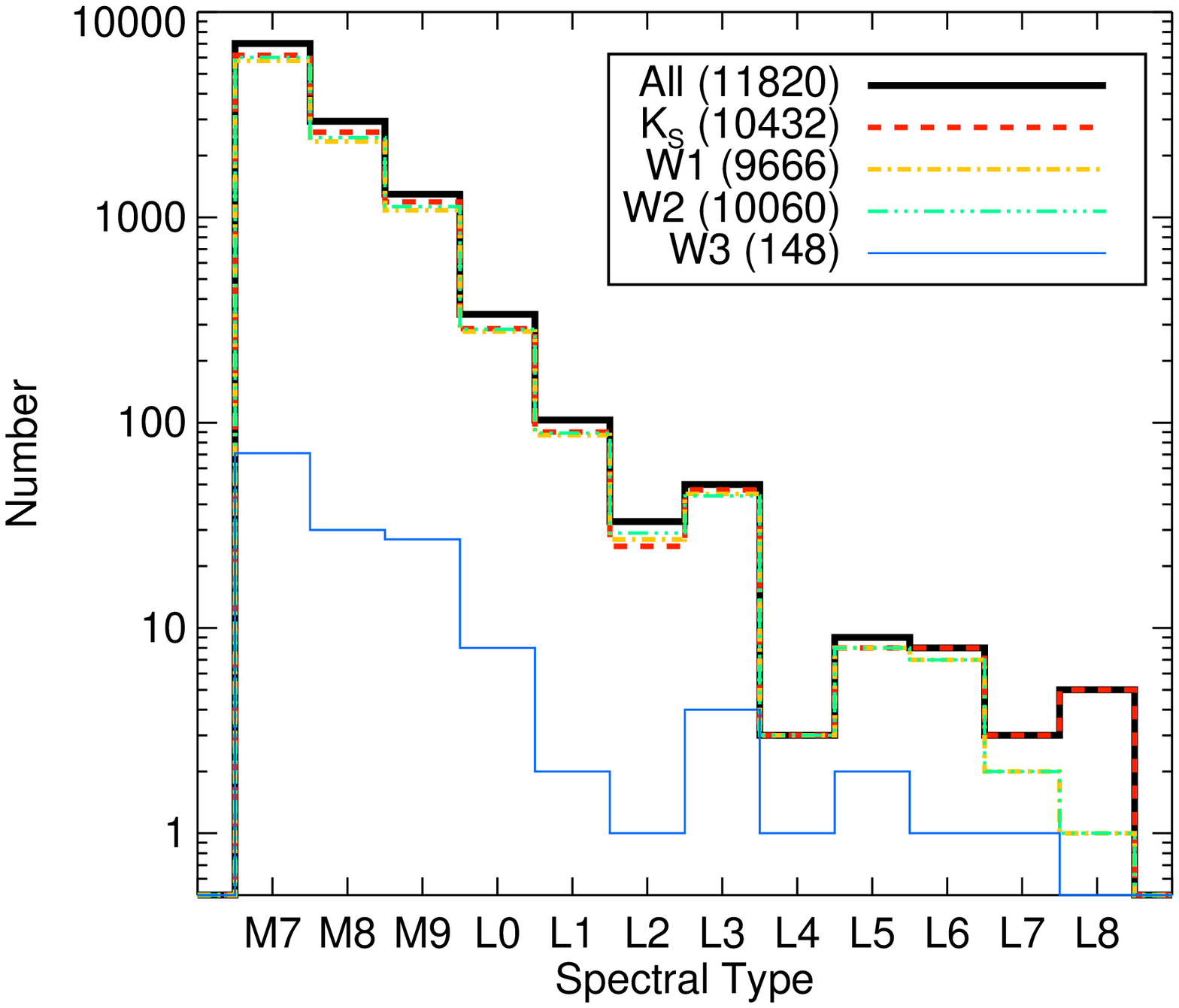} 
\caption[Number of dwarfs as a function of spectral type; number with good photometry in each band is shown.]{Number of dwarfs in the BUD sample as a function of spectral type. The entire spectroscopic sample is shown (black solid line) in addition to the number of dwarfs with good photometry (as defined by the flag and uncertainty cuts described in Section~\ref{sec:phot}) in each of the SDSS, 2MASS, and WISE bands (detailed in legend). The left panel shows the $i$-, $z$-, $J$-, and $H$-bands while the right panel shows the $K_S$-, $W1$-, $W2$- and $W3$-bands; the bands are separated only for clarity. There are only five ultracool dwarfs with good $W4$ photometry, so they are not included in the figure.} \label{fig:sthist}
\end{figure}

The BUD sample consists of spectroscopic data of late-M and L dwarfs from S10, W11, and the new DR10 BOSS observations discussed in Section~\ref{sec:sample_BUD}. Some overlap exists between the samples as BOSS re-observed some targets with the goal of obtaining higher quality spectra. For each target with multiple observations, we choose the spectrum with the highest S/N; our final sample include 8753 M7-M9 dwarfs from W11, 370 L dwarfs from S10, 2516 M7-M9 dwarfs from DR10, and 181 L dwarfs from DR10. The spectral type distribution of the 11820 ultracool dwarfs in the BUD sample is shown in Figure~\ref{fig:sthist}. The number of objects steadily declines because both the intrinsic luminosities and the space densities of L dwarfs decline with later spectral type \citep[e.g.,][]{Cruz2007}. The BUD sample has 28 dwarfs with spectral types L4-L8, compared to 523 L0-L3 dwarfs and 11,269 M7-M9 dwarfs. 

\subsection{SDSS Photometry}
While the S10 and W11 samples already include DR7 SDSS photometry \citep[in the $ugriz$ filters;][]{Fukugita1996}, we retrieved photometry for the entire BUD sample from the DR10 database. Compared to DR7, DR10 includes an additional 5200~deg$^{2}$ of SDSS imaging and a complete reprocessing of the original imaging \citep[initially included in DR8;][]{Aihara2011} in addition to improved astrometry \citep[first included in DR9;][]{Ahn2012}. Because our objects are primarily faint, the SDSS magnitudes for the BUD sample were calculated on the asinh system \citep{Lupton1999}. We obtained photometry based on an object ID match in the DR10 database; each spectroscopic observation is linked to a unique photometric object. Of the 11820 dwarfs in the DR10 BUD sample, 56 do not have matches in the DR10 photometry. We include photometry from DR7 for those dwarfs.

Table~\ref{tab:flag} lists the flags that we used to select good SDSS photometry. Based on the description of the ``CLEAN" flag for DR10 photometry\footnote{\url{http://www.sdss3.org/dr10/algorithms/photo\_flags\_recommend.php}}, we exclude objects that are saturated (SATURATED flag), blended with other objects but lacking deblended photometry (NODEBLEND), too near the border of a field to be properly reduced (EDGE), have poorly selected peaks (PEAKCENTER, NOTCHECKED, DEBLEND\_NOPEAK) or interpolation errors (PSF\_FLUX\_INTERP, BAD\_COUNTS\_ERROR, or INTERP\_CENTER). Based on these flag cuts, a total of 723 dwarfs have inaccurate photometry in either the $i$- or the $z$-band. 

\begin{deluxetable}{lrr} \tablewidth{0pt} \tabletypesize{\scriptsize}
\tablecaption{Flag Cuts on SDSS Photometry \label{tab:flag} }
\tablehead{ \colhead{Flag}  & \multicolumn{2}{c}{\# with flag set} \\
\colhead{} &  \colhead{$i$} &  \colhead{$z$} } 
\startdata
SATURATED             & 9 & 5 \\
EDGE\tablenotemark{a}              & 52 & 39 \\
NODEBLEND\tablenotemark{b}         & 334 & 334 \\
PEAKCENTER        & 45 & 60 \\
NOTCHECKED\tablenotemark{a}       & 70 & 34 \\
DEBLEND\_NOPEAK    & 3 & 5 \\
PSF\_FLUX\_INTERP\tablenotemark{c}   & 70 & 95 \\
BAD\_COUNTS\_ERROR  & 2 & 0 \\
INTERP\_CENTER\tablenotemark{c}     & 90 & 122 \\
\hline
all flags combined & 545 & 560 
\enddata
\tablenotetext{a}{EDGE and NOTCHECKED are usually, but not always, set for the same objects.}
\tablenotetext{b}{These are the same 334 objects because the NODEBLEND flag is triggered in $r$, then set in all photometric bands.}
\tablenotetext{c}{PSF\_FLUX\_INTERP and INTERP\_CENTER are usually, but not always, set for the same objects.}
\end{deluxetable}

In addition to selecting a clean photometric sample based on flags, we also required low uncertainties in each band of photometry that was included in the calculation of median colors and the color locus. We selected uncertainty cuts for each band by examining the error distribution. Each distribution was first fit by a Gaussian, then the uncertainty cut was selected as the mean of the Gaussian plus two times the dispersion, which falls near 0.03 for both $i$ and $z$. The uncertainty cuts are given in Table~\ref{tab:unc} and the spectral type distributions of objects meeting both the flag and uncertainty criteria in each band are shown in Figure~\ref{fig:sthist}.

\begin{deluxetable}{ll} \tablewidth{0pt} \tabletypesize{\scriptsize}
\tablecaption{Uncertainty Limits \label{tab:unc} }
\tablehead{ \colhead{Band}  & \colhead{$\sigma$ limit} } 
\startdata
$i$ & 0.029  \\
$z$ & 0.035  \\
$J$ & 0.176 \\
$H$ & 0.218  \\
$K_S$ & 0.264  \\
$W1$ & 0.042  \\
$W2$ & 0.103  \\
$W3$ & 0.200  \\
$W4$ & 0.300  
\enddata
\end{deluxetable}

The photometry is not corrected for extinction due to the proximity of the BOSS ultracool dwarfs to the Sun; the median distance of the BUD sample is $\sim$100~pc. The extinction corrections typically used by SDSS are calculated for extragalactic objects and so include all Galactic dust \citep{Schlegel1998}. While the scale height of the dust disk has been measured at 120--150~pc \citep{Kalberla2009}, data from local M dwarfs \citep{Jones2011} suggest that the Sun is located in a bubble with a radius of 150~pc and a dust density of 40\% of normal extinction values. Based on the complex local structure of dust, the extinction of the BUD sample varies significantly with both distance and position. Thus, applying the full SDSS extinction corrections to the entire sample would artificially de-redden the majority of the BUD objects.

While the SDSS matching is based on the catalog IDs, the matching to 2MASS and WISE is based on a coordinate cross-match. To investigate the contamination of the coordinate cross-match, we selected a random subsample of the BUD sample and queried for all primary objects within 5\arcsec. Of the $\sim1500$ objects queried, 4.3\% had two matches within 5\arcsec, and 0.1\% had three matches. Within those 4.4\%, the majority of the nearby objects were both bluer and fainter, and therefore less likely to be above the detection limits of 2MASS and WISE than the ultracool dwarf. Only 1\% were either brighter or redder. The contamination rate is likely reduced by the review of objects with multiple matches in 2MASS and WISE (as described below) so the overall confusion rate among the three surveys should be less than 1\%. 

\subsection{2MASS Photometry}
We obtained photometry from 2MASS based on matching to the closest source within 5\arcsec of the SDSS coordinates, which returned 11,199 matches. Twelve of these matches were duplicates (two matches within 5\arcsec). We inspected the image for each of the duplicates and found that in each case the closer source was the better match and rejected the duplicates. The 633 dwarfs that did not match 2MASS sources are significantly fainter than the sources which returned matches; the lack of matches is likely due to the sensitivity limit of 2MASS. Flag cuts are performed on each band individually (instead of rejecting all three bands if one was poor) to include the largest possible sample of good photometry. We require each band to have reliable photometry (ph\_qual = ABCD), contain no saturated pixels (rd\_flg = 2 ), be either unblended or be properly deblended (bl\_flg$>0$), and be uncontaminated by artifacts (cc\_flg=0). Again, we selected uncertainty limits (given in Table~\ref{tab:unc}) based on the mean and sigma of a Gaussian fit to the uncertainty distribution; for the $JHK_S$ bands, these values are close to 0.2. The number and spectral type distribution of sources with photometry that passed quality cuts in each band are shown in Figure~\ref{fig:sthist}. 

\subsection{WISE Photometry} 
WISE photometry was obtained from the ALLWISE catalog based on a match to the closest source within 5\arcsec~of the SDSS coordinates, returning a total of 11,689 matches, 17 of which were duplicates. We reviewed each duplicate match and found again that the closer match was always the better match. As a check on our procedure, we compared the 2MASS magnitudes associated with WISE to those from our 2MASS cross-match. All but seven sources (0.05\% of the total) had identical magnitudes in both of our selected samples. We inspected images for those sources, finding that the WISE photometry for those sources was a blend of two objects resolved in SDSS and 2MASS, but unresolved in WISE due to the large point spread function of the WISE photometry \citep[6\arcsec;][]{Wright2010}. We rejected the WISE data for the seven sources with blended photometry.

Flag cuts were performed on each band. We required each band to be marked as reliable photometry (ph\_qual=ABC), uncontaminated (cc\_flags = 0), not part of an extended source (ext\_flg$<$2), relatively uncontaminated by the moon (moon\_lev $<$5) and less than 20\% saturated. The 329 dwarfs not found in the WISE catalog have no clear bias in color or magnitude, but a review of SDSS images shows that objects not found in WISE are likely to have an additional point source within $\sim$5-10\arcsec. These objects could contribute to blending effects in the WISE data. We selected our uncertainty limits in the $W1$ and $W2$ band by again fitting a Gaussian function to the error distribution (limits shown in Table~\ref{tab:unc}). The $W3$- and $W4$-band uncertainty distributions peaked at values $>$0.2 mag, so we selected limits based on those of the $H$-, $K$- and $W1$-bands. 

After applying the magnitude and uncertainty cuts, the resulting numbers of ultracool dwarfs with good photometry in $W3$ and $W4$ are quite low (148 and 5 respectively) compared to the numbers of dwarfs with $W1$ and $W2$ photometry ($\sim$9800). The lack of $W3$ and $W4$ photometry for the majority of the BUD sample is due primarily to the depth of the WISE survey in those bands (average 5$\sigma$ sensitivity limits of $W3<11.5$ and $W4<8.1$ compared to $W3<16.8$ and $W4<16.0$)\footnote{from the ALLWISE explanatory supplement \url{http://wise2.ipac.caltech.edu/docs/release/allwise/expsup/sec2\_3a.html}}. The spectral type distributions of dwarfs with good photometry in $W1$, $W2$, and $W3$ are shown in Figure~\ref{fig:sthist}.

Due to the orbital pattern of the WISE satellite, the number of exposures on each region of the sky varied from 1 to 3000. The point source catalog includes a variable flag in each band to alert users to any objects showing variability. Only one object in our catalog shows significant variability (var\_flg$>5$ in $W1$ and $W2$) while passing our photometric quality cuts, an M9 dwarf (SDSS J155057.6+401255.2) first identified by W11. We do not exclude the WISE photometry for this object because the mean values for $W1$ and $W2$ were not peculiar. 

\section{Colors}
\label{sec:colors}
Because the BUD sample was selected using an $i-z$ color cut intended to include photometry outliers in those bands, the associated photometry is free from most biases. Below, we examine the SDSS/2MASS/WISE colors of the BUD sample, both with respect to spectral type and in terms of the ultracool dwarf color locus. 

\subsection{Correlation of Colors With Spectral Type}
\label{sec:medcol}
\begin{figure*}
\includegraphics[width=0.95\linewidth]{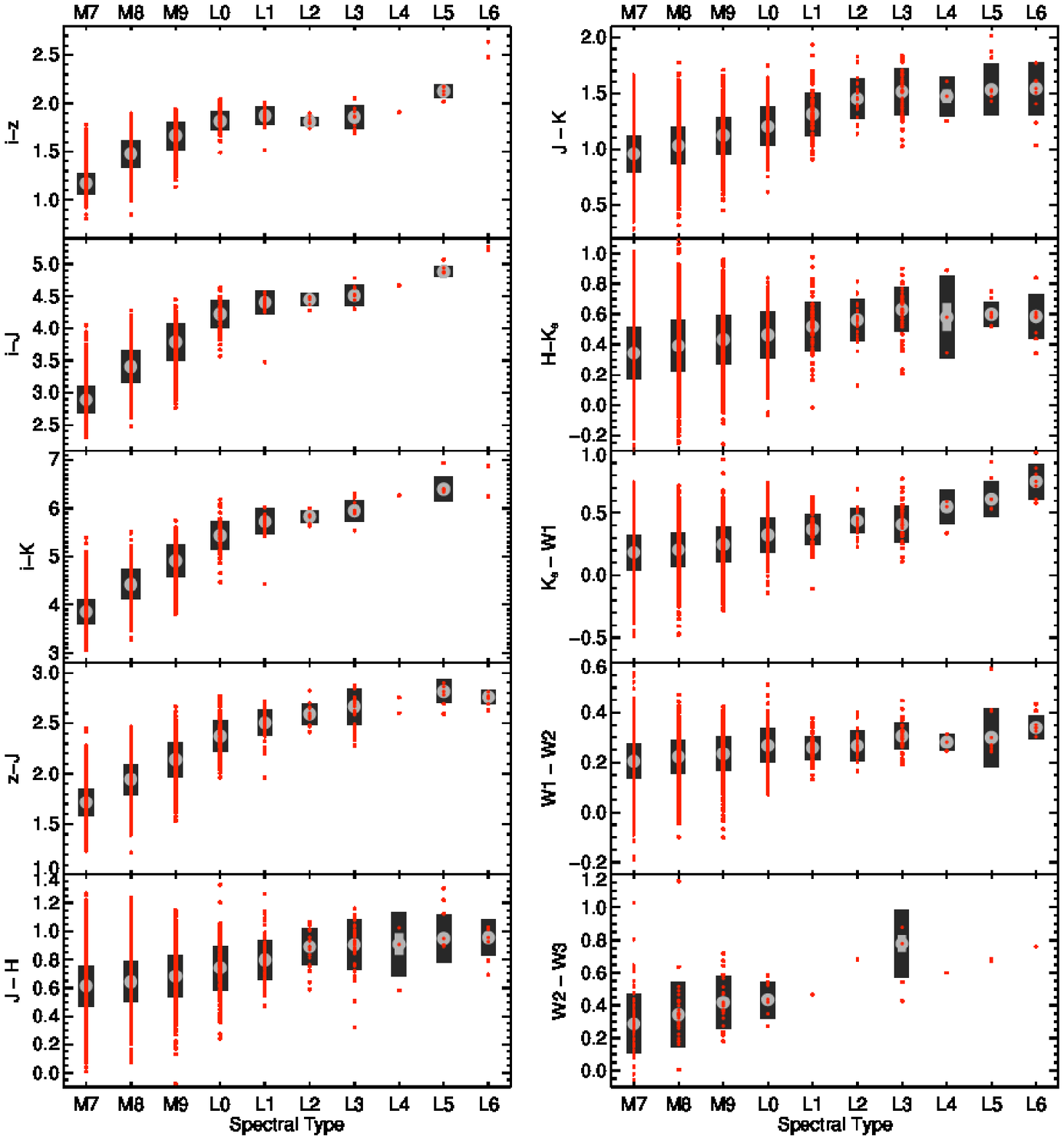} 
\caption[Eight different SDSS-2MASS-WISE colors as a function of spectral type.]{Median colors as a function of spectral type for the BUD sample. In each panel, individual objects are shown (small red circles) in addition to the median (large light grey circles) and standard deviations (black bars). The median and standard deviation values are given in Table~\ref{tab:medcol} and discussed in Section~\ref{sec:medcol}} \label{fig:medcol}
\end{figure*}

The correlation of color with spectral type is particularly useful when selecting objects from photometric surveys that warrant further investigation, and is essential in the identification of peculiar objects. The median $i-z$, $i-J$, $z-J$, $J-H$, $H-K_S$, $K_S-W1$, $W1-W2$, and $W2-W3$ colors as a function of optical spectral type are given in Table~\ref{tab:medcol} and shown in Figure~\ref{fig:medcol}. Each median is calculated using only good photometry (as defined in Section~\ref{sec:phot}). Medians are listed and shown for spectral type bins where there are at least three dwarfs with good photometry in both bands. The $\sigma$ listed in Table~\ref{tab:medcol} and shown in Figure~\ref{fig:medcol} is the standard deviation, which reflects the intrinsic spread in colors rather than the uncertainty in the median value. 

\begin{deluxetable*}{l | rll | rll | rll | rll | rll} \tablewidth{0pt} \tabletypesize{\footnotesize} 
\tablecaption{Median Colors of the BUD Sample\label{tab:medcol} }
\tablehead{
\colhead{ST} & \multicolumn{3}{c}{$i-z$} & \multicolumn{3}{c}{$i-J$} &\multicolumn{3}{c}{$i-K_S$} & \multicolumn{3}{c}{$z-J$} & \multicolumn{3}{c}{$J-H$} \\
\colhead{}  & \colhead{\#} & \colhead{med} &\colhead{$\sigma$} & \colhead{\#} & \colhead{med} &\colhead{$\sigma$}& \colhead{\#} & \colhead{med} &\colhead{$\sigma$}& \colhead{\#} & \colhead{med} &\colhead{$\sigma$} & \colhead{\#} & \colhead{med} &\colhead{$\sigma$}}
\startdata
M7 & 5899 & 1.17 & 0.11 & 5898 & 2.89 & 0.21 & 5557 & 3.86 & 0.25 & 6005 & 1.72 & 0.13 & 6316 & 0.61 & 0.14 \\
M8 & 1769 & 1.48 & 0.14 & 1795 & 3.41 & 0.26 & 1770 & 4.42 & 0.31 & 2172 & 1.95 & 0.15 & 2519 & 0.64 & 0.15 \\
M9 & 581 & 1.66 & 0.14 & 600 & 3.79 & 0.29 & 589 & 4.91 & 0.35 & 916 & 2.14 & 0.17 & 1147 & 0.68 & 0.15 \\
L0 & 76 & 1.82 & 0.10 & 76 & 4.22 & 0.21 & 77 & 5.44 & 0.29 & 153 & 2.37 & 0.15 & 274 & 0.74 & 0.16 \\
L1 & 30 & 1.87 & 0.09 & 30 & 4.41 & 0.18 & 30 & 5.73 & 0.26 & 65 & 2.50 & 0.13 & 88 & 0.80 & 0.14 \\
L2 & 8 & 1.81 & 0.05 & 7 & 4.45 & 0.10 & 7 & 5.83 & 0.12 & 14 & 2.59 & 0.10 & 25 & 0.89 & 0.13 \\
L3 & 9 & 1.86 & 0.12 & 9 & 4.51 & 0.16 & 9 & 5.95 & 0.22 & 26 & 2.67 & 0.18 & 46 & 0.91 & 0.18 \\
L4 & 0 & \nodata & 0.00 & 0 & \nodata & 0.00 & 0 & \nodata & 0.00 & 0 & \nodata & 0.00 & 3 & 0.91 & 0.23 \\
L5 & 5 & 2.13 & 0.06 & 5 & 4.88 & 0.09 & 5 & 6.40 & 0.25 & 6 & 2.82 & 0.11 & 7 & 0.95 & 0.17 \\
L6 & 0 & \nodata & 0.00 & 0 & \nodata & 0.00 & 0 & \nodata & 0.00 & 6 & 2.76 & 0.07 & 8 & 0.96 & 0.13 \\
\hline
\hline
ST & \multicolumn{3}{c}{$J-K_S$} & \multicolumn{3}{c}{$H-K_S$} & \multicolumn{3}{c}{$K_S-W1$} & \multicolumn{3}{c}{$W1-W2$} & \multicolumn{3}{c}{$W2-W3$} \\
  & \# & med & $\sigma$ & \# & med & $\sigma$ & \# & med & $\sigma$ & \# & med & $\sigma$ & \# & med & $\sigma$ \\
\hline
M7 & 6072 & 0.96 & 0.17 & 5996 & 0.34 & 0.17 & 5377 & 0.18 & 0.14 & 5599 & 0.20 & 0.07 & 62 & 0.29 & 0.18 \\
M8 & 2516 & 1.03 & 0.17 & 2482 & 0.39 & 0.17 & 2220 & 0.20 & 0.14 & 2260 & 0.22 & 0.07 & 28 & 0.34 & 0.20 \\
M9 & 1145 & 1.12 & 0.17 & 1150 & 0.43 & 0.16 & 1044 & 0.25 & 0.14 & 1047 & 0.24 & 0.07 & 22 & 0.42 & 0.16 \\
L0 & 273 & 1.20 & 0.17 & 276 & 0.46 & 0.16 & 258 & 0.32 & 0.14 & 271 & 0.27 & 0.07 & 7 & 0.44 & 0.11 \\
L1 & 87 & 1.31 & 0.19 & 90 & 0.52 & 0.16 & 83 & 0.37 & 0.12 & 82 & 0.26 & 0.05 & 0 & \nodata & 0.00 \\
L2 & 23 & 1.45 & 0.17 & 25 & 0.56 & 0.14 & 22 & 0.44 & 0.10 & 26 & 0.27 & 0.06 & 0 & \nodata & 0.00 \\
L3 & 46 & 1.52 & 0.21 & 47 & 0.63 & 0.15 & 45 & 0.41 & 0.15 & 43 & 0.31 & 0.05 & 4 & 0.78 & 0.21 \\
L4 & 3 & 1.47 & 0.18 & 3 & 0.58 & 0.27 & 3 & 0.55 & 0.14 & 3 & 0.28 & 0.03 & 0 & \nodata & 0.00 \\
L5 & 7 & 1.53 & 0.23 & 8 & 0.60 & 0.08 & 7 & 0.61 & 0.14 & 8 & 0.30 & 0.12 & 0 & \nodata & 0.00 \\
L6 & 8 & 1.54 & 0.23 & 8 & 0.59 & 0.15 & 7 & 0.75 & 0.14 & 7 & 0.34 & 0.05 & 0 & \nodata & 0.00
\enddata
\end{deluxetable*}

The SDSS/2MASS median colors show good agreement with the colors given in W11 and S10. The $J-H$ and $H-K_S$ colors have larger dispersions than those of W11, perhaps due to the stricter limits on uncertainty applied in that study ($<$0.05 mag). The $i-z$, $i-J$, $i-K_S$ and $z-J$ colors are a strong function of spectral type for M7-L1 dwarfs, each color tracing the same changing spectral slope from 7000~\AA~to 12000~\AA. For L1 to L6 dwarfs, $i-z$ remains constant, as both the $i$ and $z$ bands are affected by absorption from the dramatic broadening of the 7665/7699~\AA~K I doublet. The $i-J$, $i-K_S$ and $z-J$ colors are a weak function of spectral type for L1-L5 dwarfs, tracing the ratio of the K I doublet absorption to the flux at the $J$-band peak. 

While the $i-z$, $i-J$, and $z-J$ colors have relatively tight dispersions compared to the large color range, the $J-H$, $J-K_S$, and $H-K_S$ colors have larger dispersions and smaller changes with respect to spectral type. For late-M and L dwarfs, these colors show much stronger correlations with age and cloud properties \citep[e.g.,][]{Burgasser2008a,Cruz2009,Faherty2009,Schmidt2010a} than spectral type. As our sample was not selected based on 2MASS colors, these $J-H$ and $H-K_S$ colors should follow the true median with respect to spectral type, rather than a redder sample selected to avoid earlier-type, bluer objects (similar to S10). 

Generally, our 2MASS-WISE colors are consistent with those of \citet{Kirkpatrick2011}. The $K_S-$W1 color shows a slight increase with spectral type from M7 to L3, and a more dramatic increase for later-L dwarfs. The pattern is similar to $K_S-W2$ as a function of spectral type, used by \citet{Kirkpatrick2011} to distinguish between L and T dwarfs. The $W1-W2$ color increases slightly with spectral type over the M7-L6 range, with variations in the color likely due to the relatively small number of dwarfs with good WISE photometry in the mid-L dwarf spectral type bins. The $W2-W3$ color is remarkable in the fraction of red outliers; six (4\% of the 136 meeting quality cuts) of the dwarfs with good photometry have $W2-W3>1.5$, well above 3$\sigma$ from the mean for each spectral type. We exclude these six outliers with $W2-W3 > 1.5$ from our calculated median colors; the outliers are discussed below. These corrected $W2-W3$ colors also show a slight increase with spectral type from M7-L6.

Over the M7-L6 spectral range, there is no one set of colors that can be used to estimate spectral type. The M7-L0 dwarfs can be distinguished with the $i-z$, $i-J$, $i-K_S$, and $z-J$ colors, but the L1-L3 spectral types only show a weak correlation with the $i-J$ through $H-K_S$ colors. The BUD color data for L4 and later types are sparse, but the $i-z$, $i-J$, and $K_S-W1$ colors are all relatively strong functions of late-L spectral type \citep[S10,][]{Kirkpatrick2011}. Given the changing relationships between color and spectral type, ultracool dwarfs can only be fully spectral typed either with photometry spanning much of the SDSS/2MASS/WISE bands or with spectroscopic data. 

Six ultracool dwarfs with good WISE photometry were found to have $W2-W3 > 1.5$, representing a deviation from the median $W2-W3$ color of $>3\sigma$. Inspection of the images for these six dwarfs reveals that four are likely to be blends in the WISE bands due to nearby objects resolved in SDSS and 2MASS but unresolved in WISE. Two remaining sources (M7 SDSS J053503.56$-$001511.7 and L0 SDSS J233358.42+005012.1) have no obvious contaminants so may have real infrared color excesses. \citet{Theissen2014} provide a detailed examination of M dwarfs from W11 with WISE color excesses, but do not recover SDSS J053503.56$-$001511.7 because its SDSS photometry is contaminated by a diffraction spike (thus excluding it from the sample examined). 

As discussed by \citet{Theissen2014}, infrared color excesses on M dwarfs in the WISE bands are more likely the result of a debris disk surrounding the star \citep{Carpenter2009} than an accretion disk. Another possibility is an unresolved faint ultracool (likely T dwarf) companion \citep[e.g., the M8.5/T6 binary SRC 18450$-$6357 recovered in the debris disk search of][]{Avenhaus2012}. If these sources do have disks, their detections in the WISE passbands (located at relatively short wavelengths for cool disks) indicate that they are likely to be warm disks \citep{Heng2013}.

\subsection{SDSS--2MASS--WISE Ultracool Dwarf Locus}
\label{sec:loci} 
The SDSS--2MASS stellar locus, defined by \citet{Covey2007} using $ugrizJHK_S$ photometry, is important both for characterizing the colors of main sequence stars and for identifying peculiar or non-stellar objects due to their distances from the colors defined in the locus. \citet{Davenport2014} expanded the stellar locus defined in \citet{Covey2007} to cover the entire SDSS--2MASS--WISE color space. Because the stellar locus is defined in terms of $g-i$ color, ultracool dwarfs (which are typically too faint to be detected in $g$) are not included in these studies. We use the BUD sample to define an ultracool dwarf locus for M7--L3 dwarfs using the $i-J$ color.

\begin{figure*}
\includegraphics[width=0.95\linewidth]{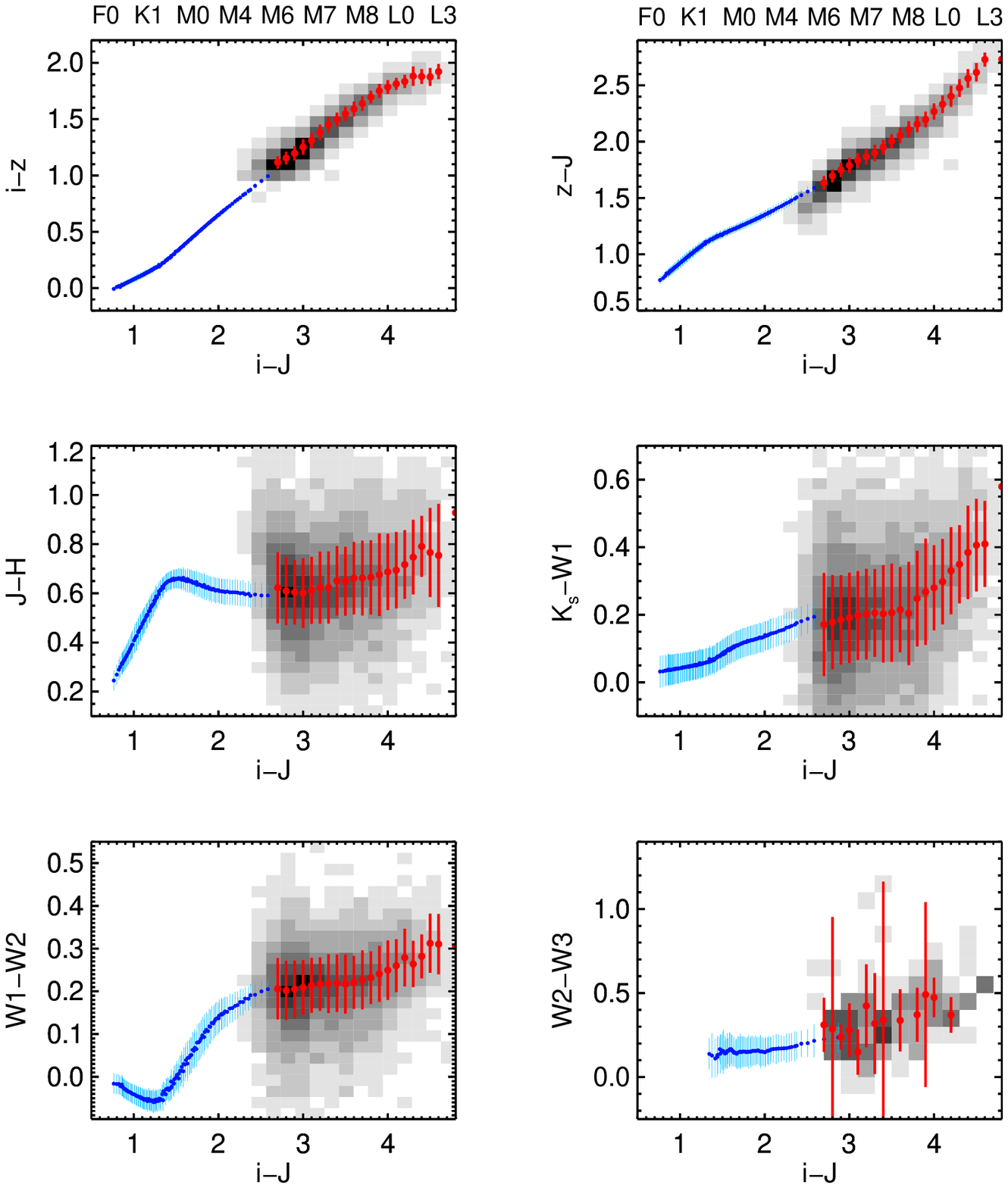} 
\caption[Six colors as a function of $i-J$ shown with the color locus for the stellar population.]{The color locus for several filter combinations defined in term of $i-J$ color. In each panel, the stellar locus from  \citet{Davenport2014} is shown (blue points) in addition to the density of BUD stars (greyscale pixels) and the mean and dispersion of the ultracool dwarf locus (red points). Offsets between the two datasets are primarily due to the small numbers in the reddest bins of the Davenport et al.\ color sample; the $K_S-W1$ color for those bins also suffers a slight bias towards redder objects due to the deeper photometric limit of $W1$ compared to $K_S$.} \label{fig:locus}
\end{figure*}

We measure the ultracool dwarf locus from the BUD sample in steps of $\delta(i-J)=0.1$ for $2.8 < i-J < 4.6$. In each color bin, we apply the uncertainty cuts described in Section~\ref{sec:medcol} and calculate the median and standard deviation of that color. The results for seven adjacent colors (from $i-z$ to $W2-W3$) are given in Table~\ref{tab:locus}, including the standard deviation of the color ($\sigma$) and the number of objects in each bin. The ultracool dwarf locus is shown compared to the stellar locus in six of the seven colors (excluding $H-K$, which has a shape similar to the $J-H$ color) in Figure~\ref{fig:locus}. For reference, rough spectral types (from F0 to L3) are also shown. 

\setlength{\tabcolsep}{3pt}
\begin{deluxetable*}{l | rll | rll | rll | rll | rll | rll | rll }\tablewidth{0pt} \tabletypesize{\small} 
\tablecaption{The SDSS--2MASS--WISE Utracool Dwarf Locus \label{tab:locus}}
\tablehead{ \multicolumn{1}{l}{$i-J$} & \multicolumn{3}{c}{$i-z$} & \multicolumn{3}{c}{$z-J$} & \multicolumn{3}{c}{$J-H$} & \multicolumn{3}{c}{$H-K_s$} & \multicolumn{3}{c}{$K_s-W1$} & \multicolumn{3}{c}{$W1-W2$} & \multicolumn{3}{c}{$W2-W3$}  \\
\colhead{ }& \colhead{\#} & \colhead{mean} & \colhead{$\sigma$}& \colhead{\#}  & \colhead{mean} & \colhead{$\sigma$}& \colhead{\#}  & \colhead{mean} & \colhead{$\sigma$}& \colhead{\#}  & \colhead{mean} & \colhead{$\sigma$}& \colhead{\#}  & \colhead{mean} & \colhead{$\sigma$}& \colhead{\#}  & \colhead{mean} & \colhead{$\sigma$}& \colhead{\#}  & \colhead{mean} & \colhead{$\sigma$} }
\startdata
2.7 & 1045 & 1.116 & 0.055 & 1070 & 1.636 & 0.059 & 1102 & 0.623 & 0.144 & 1034 & 0.331 & 0.176 & 903 & 0.171 & 0.153 & 935 & 0.206 & 0.072 & 9 & 0.311 & 0.161 \\
2.8 & 1157 & 1.154 & 0.058 & 1201 & 1.697 & 0.058 & 1256 & 0.609 & 0.139 & 1184 & 0.339 & 0.162 & 1087 & 0.178 & 0.139 & 1115 & 0.202 & 0.069 & 7 & 0.287 & 0.665 \\
2.9 & 1069 & 1.196 & 0.063 & 1109 & 1.749 & 0.064 & 1138 & 0.604 & 0.131 & 1092 & 0.337 & 0.162 & 1025 & 0.185 & 0.132 & 1049 & 0.205 & 0.067 & 8 & 0.238 & 0.075 \\
3.0 & 894 & 1.252 & 0.068 & 919 & 1.789 & 0.067 & 958 & 0.601 & 0.142 & 923 & 0.346 & 0.171 & 847 & 0.191 & 0.134 & 859 & 0.209 & 0.064 & 13 & 0.280 & 0.159 \\
3.1 & 646 & 1.310 & 0.067 & 690 & 1.833 & 0.067 & 730 & 0.611 & 0.137 & 727 & 0.348 & 0.151 & 661 & 0.198 & 0.131 & 663 & 0.214 & 0.064 & 9 & 0.149 & 0.135 \\
3.2 & 492 & 1.383 & 0.067 & 529 & 1.868 & 0.071 & 586 & 0.623 & 0.147 & 586 & 0.368 & 0.162 & 535 & 0.202 & 0.134 & 535 & 0.217 & 0.063 & 9 & 0.424 & 0.247 \\
3.3 & 463 & 1.451 & 0.068 & 557 & 1.900 & 0.072 & 612 & 0.624 & 0.146 & 613 & 0.376 & 0.172 & 548 & 0.206 & 0.130 & 549 & 0.218 & 0.069 & 10 & 0.319 & 0.300 \\
3.4 & 422 & 1.498 & 0.060 & 516 & 1.951 & 0.066 & 603 & 0.651 & 0.142 & 600 & 0.402 & 0.166 & 536 & 0.203 & 0.144 & 529 & 0.219 & 0.067 & 9 & 0.340 & 0.822 \\
3.5 & 369 & 1.547 & 0.062 & 503 & 2.004 & 0.062 & 602 & 0.649 & 0.140 & 595 & 0.399 & 0.171 & 525 & 0.206 & 0.145 & 532 & 0.217 & 0.069 & 0 & \nodata & \nodata \\
3.6 & 299 & 1.590 & 0.066 & 410 & 2.056 & 0.068 & 514 & 0.661 & 0.147 & 511 & 0.399 & 0.164 & 435 & 0.215 & 0.126 & 445 & 0.221 & 0.062 & 6 & 0.338 & 0.185 \\
3.7 & 219 & 1.638 & 0.065 & 359 & 2.109 & 0.072 & 469 & 0.662 & 0.152 & 459 & 0.430 & 0.171 & 419 & 0.204 & 0.153 & 422 & 0.227 & 0.069 & 0 & \nodata & \nodata \\
3.8 & 180 & 1.696 & 0.056 & 309 & 2.155 & 0.072 & 413 & 0.665 & 0.151 & 406 & 0.431 & 0.159 & 380 & 0.248 & 0.141 & 383 & 0.232 & 0.062 & 10 & 0.371 & 0.160 \\
3.9 & 110 & 1.750 & 0.061 & 228 & 2.197 & 0.069 & 308 & 0.675 & 0.168 & 307 & 0.445 & 0.177 & 293 & 0.268 & 0.157 & 289 & 0.241 & 0.066 & 7 & 0.491 & 0.550 \\
4.0 & 98 & 1.785 & 0.060 & 185 & 2.268 & 0.071 & 251 & 0.687 & 0.155 & 246 & 0.442 & 0.143 & 235 & 0.280 & 0.126 & 232 & 0.249 & 0.070 & 6 & 0.474 & 0.117 \\
4.1 & 71 & 1.813 & 0.052 & 146 & 2.332 & 0.077 & 213 & 0.694 & 0.155 & 211 & 0.471 & 0.143 & 186 & 0.298 & 0.126 & 183 & 0.260 & 0.062 & 0 & \nodata & \nodata \\
4.2 & 50 & 1.835 & 0.058 & 108 & 2.403 & 0.093 & 153 & 0.717 & 0.139 & 154 & 0.479 & 0.135 & 140 & 0.331 & 0.129 & 140 & 0.279 & 0.067 & 7 & 0.370 & 0.106 \\
4.3 & 38 & 1.882 & 0.086 & 69 & 2.476 & 0.079 & 113 & 0.747 & 0.152 & 116 & 0.500 & 0.131 & 110 & 0.350 & 0.115 & 103 & 0.264 & 0.054 & 0 & \nodata & \nodata \\
4.4 & 28 & 1.878 & 0.066 & 57 & 2.562 & 0.082 & 96 & 0.791 & 0.124 & 94 & 0.487 & 0.162 & 86 & 0.385 & 0.138 & 91 & 0.282 & 0.051 & 0 & \nodata & \nodata \\
4.5 & 10 & 1.875 & 0.079 & 22 & 2.615 & 0.082 & 37 & 0.766 & 0.181 & 37 & 0.612 & 0.169 & 33 & 0.406 & 0.137 & 30 & 0.312 & 0.069 & 0 & \nodata & \nodata \\
4.6 & 6 & 1.922 & 0.068 & 12 & 2.729 & 0.062 & 22 & 0.754 & 0.209 & 21 & 0.592 & 0.111 & 22 & 0.409 & 0.127 & 22 & 0.310 & 0.070 & 0 & \nodata & \nodata 
\enddata
\end{deluxetable*}

Both the $i-z$ and $z-J$ colors increase rapidly as a function of $i-J$, with the $i-z$ color showing a plateau at $i-J>4$. This behavior is not surprising, as the colors measure similar portions of the spectra. The $J-H$ color increases for $0.5 < i-J < 1.4$ (F0 to M0) stars, then shows a dip through the late-M spectral types ($2.8<i-J<3.8$). This result indicates that $J-H$ is not a good color for distinguishing late-M and early-L dwarfs from early- to mid-M stars. The $K_s-W1$ color also increases with $i-J$, with a slope that steepens at $i-J\sim 3.5$ (M8 spectral type). The $W1-W2$ color also increases for stars redder than $i-J>1.2$, with a shallower slope for M dwarfs than for any other type of star. There are only a few ultracool dwarfs with $W2-W3$ colors, but those are consistent with an increase from the M0--M6 dwarfs at $W2-W3\sim0.2$ to $W2-W3\sim0.4$ for late-M and early-L dwarfs.

Examining the ultracool dwarf color locus compared to the stellar locus for F to M dwarfs emphasizes the large spread in $J-H$, $H-K_S$, and $K_S-W1$ colors; they span the full range of color space occupied by the bluer stars. Over most wavelengths, changing color is primarily a function of effective temperature, but for ultracool dwarfs the flux in the $H$ and $K_S$ bands depends more strongly on the effects of clouds and surface gravity \citep[e.g.,][]{Burgasser2008a,Cruz2009}. To select M7-L3 dwarfs from other stars, $i-z$, $i-J$, $z-J$, and $W1-W2$ are more useful than colors including the $H$ and $K_S$ bands. For L5 and later dwarfs, 2MASS-WISE colors are better discriminants of spectral type \citep[here $K_S-$W1, but also $J-$W1 and $K_S-$W2;][]{Kirkpatrick2011}. Given the complex relationships between spectral type and color, it is important to have as many bands of photometry as possible to estimate photometric spectral types. 

\section{H$\alpha$ Data For M and L dwarfs}
\label{sec:haobs}
To investigate the chromospheres of cool and ultracool dwarfs, we combine the H$\alpha$ emission from our BUD sample with both the W11 DR7 M dwarf sample and previous spectroscopic observations of L dwarf activity from the literature. We describe our SDSS H$\alpha$ measurements (Section~\ref{sec:habud}) and the sources of the L dwarf H$\alpha$ data (Section~\ref{sec:hal}) below. 

\subsection{H$\alpha$ Emission from SDSS Spectra}
\label{sec:habud}
The H$\alpha$ EWs for the BUD sample were measured using the similar methods as W11. When radial velocities were available, the spectra were first velocity-corrected to 0 km s$^{-1}$. The equivalent widths (EW) of H$\alpha$ emission lines were measured using a range of 6557.61--6571.61~\AA~for the line and 6530--6555~\AA~and 6575--6600~\AA~for the surrounding continuum. The EW uncertainties were calculated including both the flux errors and the standard deviation of the continuum ranges. 

To classify the H$\alpha$ emission in SDSS spectra, we required a minimum S/N per 1.5~\AA~pixel in the continuum region used to calculate the EW. For the M7--M9 dwarfs we adopted the criterion of W11, requiring S/N$ > 3$ to include each spectrum in out activity classification. For L dwarfs, review of the spectra indicated that a slightly higher threshold was necessary to select active and inactive dwarfs; we adopt S/N$ > 4$ for the L dwarfs included in the activity classification. These cuts limit our BUD sample to 5699 M dwarfs and 26 L dwarfs. 

Following W11, objects are classified as active if they have (a) a measured H$\alpha$ EW greater than its uncertainty, (b) an H$\alpha$ EW greater than the detection threshold of EW = 0.75~\AA, and (c) a peak height greater than three times the standard deviation (noise) of the continuum region. Weakly active dwarfs met criteria (a) and (c), but not the 0.75~\AA~detection threshold; we classified ``maybe" active dwarfs as those meeting criteria (a) and (b) but only having a peak height greater than twice the standard deviation of the continuum region. Dwarfs were classified as inactive if they passed the S/N threshold but failed to fall into the active, weakly active, or maybe active categories. We excluded the maybe and weakly active categories from subsequent analysis. 

\begin{figure*}
\includegraphics[width=0.95\linewidth]{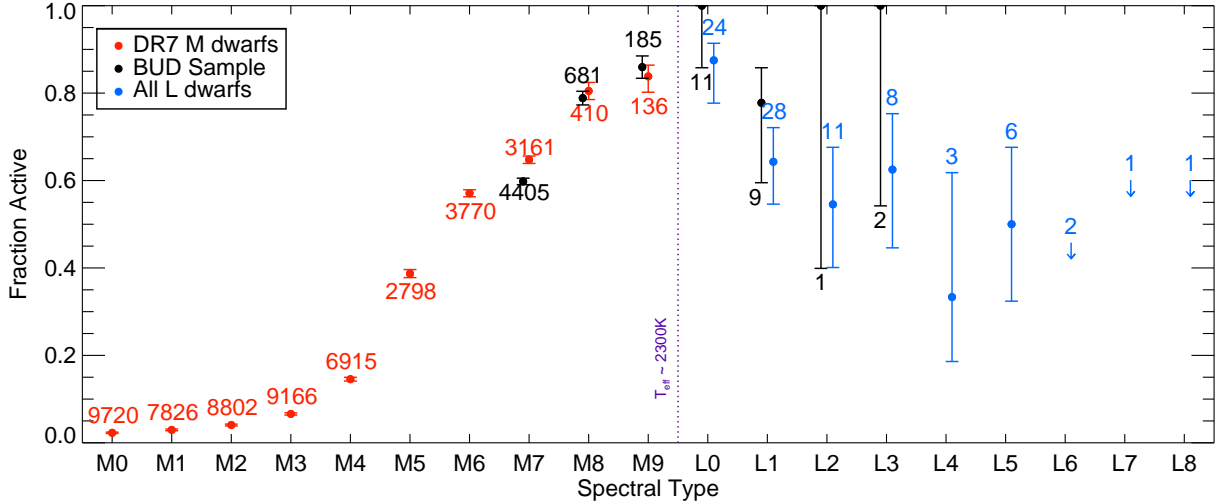} 
\caption[Fraction of active M and L dwarfs as a function of spectral type.]{Fraction of active M and L dwarfs (as determined by the presence of H$\alpha$ emission with an EW$>$0.75\AA) as a function of spectral type. Fractions computed from the W11 sample (red), from the BUD sample (black) and from the L dwarf activity sample (described in Section~\ref{sec:hal}; blue) are shown with uncertainties based on a binomial distribution. For the spectral type bins with no detections, the arrows show upper limits also based on a binomial distribution. The total number of objects used to compute the fraction (active and inactive) are shown above or below the data in corresponding colors. The \citet{Mohanty2002} threshold is shown (vertical purple dotted line) between M9 and L0, corresponding to T$_{\rm eff}$ = 2300~K (as discussed in Section~\ref{sec:disc}).} \label{fig:frac}
\end{figure*}

The fraction of active dwarfs from the BUD and W11 DR7 M dwarf samples are shown in Figure~\ref{fig:frac}. The fractions for the M8 and M9 dwarfs are the same in the two studies, but the M7 fraction differs, possibly due to the slightly different regions used to measure S/N and line flux between this work and W11. The combined sample shown here indicates a rise in activity fraction from 2\% at M0 to 88\% at L0. While the rise from early- to mid-M dwarfs has been repeatedly documented \citep[e.g.,][]{Hawley1996,Gizis2000,West2004}, the peak at L0 and the persistence of a high activity fraction ($>50$\% through L3) is a new result. The detection of the peak at a later spectral type than previously thought is made possible by the combination of high S/N early-L spectra in the BUD sample with previous results. 

The strength of activity is often quantified by the ratio of luminosity in the H$\alpha$ line to bolometric luminosity, or $L_{\rm H\alpha}/L_{\rm bol}$ \citep[e.g.,][]{Hawley1996}. This ratio removes the dependence of the measured EW on the surrounding continuum allowing comparison of chromospheric emission across a range of spectral type and T$_{\rm eff}$. The $L_{\rm H\alpha}/L_{\rm bol}$ values for M and L dwarfs are usually calculated by multiplying their H$\alpha$ EW by the ``$\chi$'' factor \citep[e.g.,][]{Walkowicz2004,West2008a}. The $\chi$ factor is calculated from the ratio of the continuum luminosity near H$\alpha$ to the bolometric luminosity; it is typically based on a handful of stars where the bolometric luminosity can be reliably estimated, then fit as a function of color or spectral type. 

\begin{figure*}
\includegraphics[width=\linewidth]{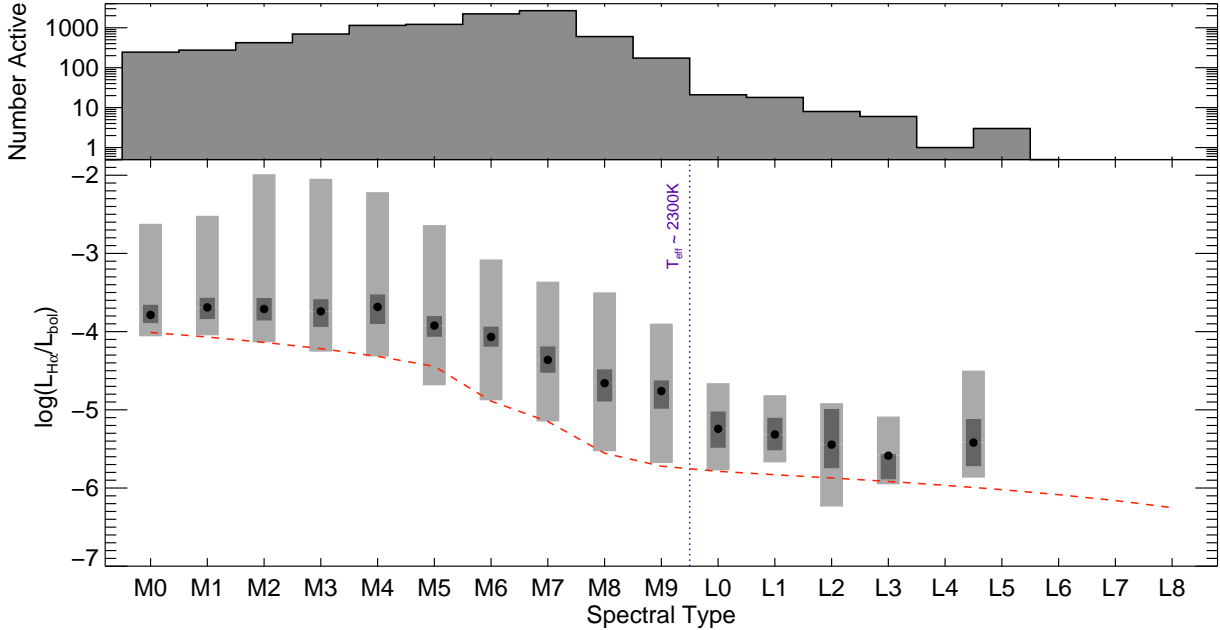} 
\caption[Activity strength as a function of spectral type.]{Top panel: Number of active dwarfs as a function of spectral type. Bottom panel: Activity strength as a function of spectral type for DR7 (M0-M6), BUD (M7-M9) and the L dwarf activity sample (L0-L5; including the BUD L dwarfs; discussed in Section~\ref{sec:hal}). The minimum and maximum (light grey bars), the interquartile range (dark grey bars) and median values (black circles) are shown. The upper envelope of the M0--M6 spectral type bins is likely affected by the serendipitous detection of H$\alpha$ emission during flares, an effect that is most noticeable for M2--M4 dwarfs where strong flares are common. In these bins, the strongest detections overestimate the H$\alpha$ range observed in quiescence. The dashed red line shows the effective lower limit of detection for SDSS data (based on the 0.75\AA~H$\alpha$ EW threshold; one L2 dwarf from the non-SDSS portion of the L activity sample falls below the limit). The \citet{Mohanty2002} threshold is also shown (vertical purple dashed line), see Section~\ref{sec:disc}.} \label{fig:lha_all}
\end{figure*}

The activity strength for the W11 DR7 M0--M6 dwarfs was calculated based on their measured EW and the $\chi$-spectral type relation from \citet{West2008a}, and for the M7--M9 dwarfs in the BUD sample using the relation from \citet{Schmidt2014a}. Activity strength as a function of spectral type for the DR7 M dwarfs (M0--M6) and the BUD sample (M7--M9) is shown in Figure~\ref{fig:lha_all}. Median activity strength is constant from M0--M4 at log($L_{\rm H\alpha}/L_{\rm bol}$) = $-3.8$, then displays a steady decline through the rest of the M spectral sequence. The L dwarf data presented in Figure~\ref{fig:lha_all} are described in Section~\ref{sec:hal}. 

When considering the observed M dwarf activity fraction and activity strengths, it is important to consider two observational biases. The first bias is due to the limits on observing H$\alpha$ emission in contrast with the photospheric emission. In Figure~\ref{fig:lha_all}, we show the EW H$\alpha > 0.75$~\AA~limit compared to the median and range of activity strengths measured from the spectra. For early-M dwarfs, the median is close to the limit, suggesting a population of active dwarfs with activity strength below the detection limit (e.g., the ``weakly active" M dwarfs discussed above). Higher resolution spectroscopy indicates that early-M dwarfs may indeed be active with lower EW \citep[e.g.,][]{Walkowicz2009}; some weakly active M dwarfs even possess H$\alpha$ in absorption \citep[e.g.,][]{Stauffer1986}. This effect on the activity fraction is mitigated by the exclusion of ``weakly active" M dwarfs, but there may be additional weakly active M dwarfs that fall below the detection threshold at the SDSS resolution and sensitivity. These objects would be included in our sample as inactive M dwarfs. 

The second bias is due to the changing stellar populations included in the SDSS M dwarf samples. The early-M dwarfs in the W11 DR7 sample are found, on average, much farther from the Galactic plane (and thus are representative of an older population) than the mid- to late-M dwarfs. When we consider M dwarfs only within 100~pc of the Galactic plane, the activity fraction shown in Figure~\ref{fig:frac} increases to 10\% at type M0 and 50\% at type M4, while the active fraction at later types remains unchanged \citep{West2008}. Preliminary examination of the BUD sample indicates similar age and activity effects for M8 and later dwarfs; the effect of the changing Galactic population on the activity fraction (and the interaction between age and activity for ultracool dwarfs) will be examined in subsequent papers on the BUD sample. 

\subsection{H$\alpha$ emission in L dwarfs}
\label{sec:hal}
Of the 551 L dwarfs in the BUD sample, 26 have sufficient S/N to classify them as either active or inactive (21 active, 3 inactive, and 2 maybe active). To supplement the sample of L dwarfs with activity classifications, we include H$\alpha$ detections and non-detections for objects that have been published as part of discovery papers \citep[e.g.,][]{Kirkpatrick1999,Kirkpatrick2000}, activity surveys \citep[e.g.,][]{Schmidt2007,Reiners2008}, and serendipitous detections \citep[e.g.,][]{Hall2002,Liebert2003,Burgasser2011b}. Data from these sources are listed in Table~\ref{tab:Lha} (we refer to this sample as the ``L dwarf activity sample"). The L dwarf activity sample includes data for active and inactive (upper limit of H$\alpha$ EW$>0.75$~\AA) BUD L dwarfs but excludes objects which did not meet the S/N$>3$ criterion.

Of the 181 L dwarfs with reported H$\alpha$ detections and upper limits, 38 were observed more than once. For those with multiple observations, we include all entries in Table~\ref{tab:Lha}, but only use one of the detections in the final L dwarf activity sample. Twelve L dwarfs are assigned upper limits from two sources; the smaller upper limit is used because it places a stronger constraint on the maximum possible emission. Eleven L dwarfs have both a detection and an upper limit from different sources and we use the detection in the L dwarf activity sample. Fifteen L dwarfs have two detections of H$\alpha$ in emission. Where the detections differ in activity strength, we use the lower detection because our goal is to examine quiescent emission and some higher detections may have been made during flares. 

The multiple detections can also be used to roughly characterize variability. The twelve L dwarfs with multiple upper limits and no detections are classified as non-variable because their emission did not become sufficiently strong for detection. Similarly, the five L dwarfs with an upper limit higher than a reported detection show no evidence of variability. For the 21 L dwarfs with multiple detections or an upper limit lower than a reported detection, we calculate both the normalized standard deviation ($\sigma_{\rm H\alpha}/\langle{\rm H\alpha}\rangle$) and the fractional variability (the total range of EW H$\alpha$ divided by the minimum EW H$\alpha$). We classify the fifteen L dwarfs with fractional variability more than one as variable L dwarfs (as noted in Table~\ref{tab:Lha}). 

Of all L dwarfs observed multiple times, 39\% are variable according to this definition. If we only include active L dwarfs with more than one H$\alpha$ detection, 47\% of those are variable. While these numbers only represent a rough estimate of the fraction of L dwarfs showing significant variability, it is a smaller then the $\sim$60\% of active M dwarfs that have been found variable \citep{Lee2010,Bell2012}. The strength of variability and the fraction of variable L dwarfs shows no dependence on activity strength or on spectral type. 

\begin{figure}
\includegraphics[width=\linewidth]{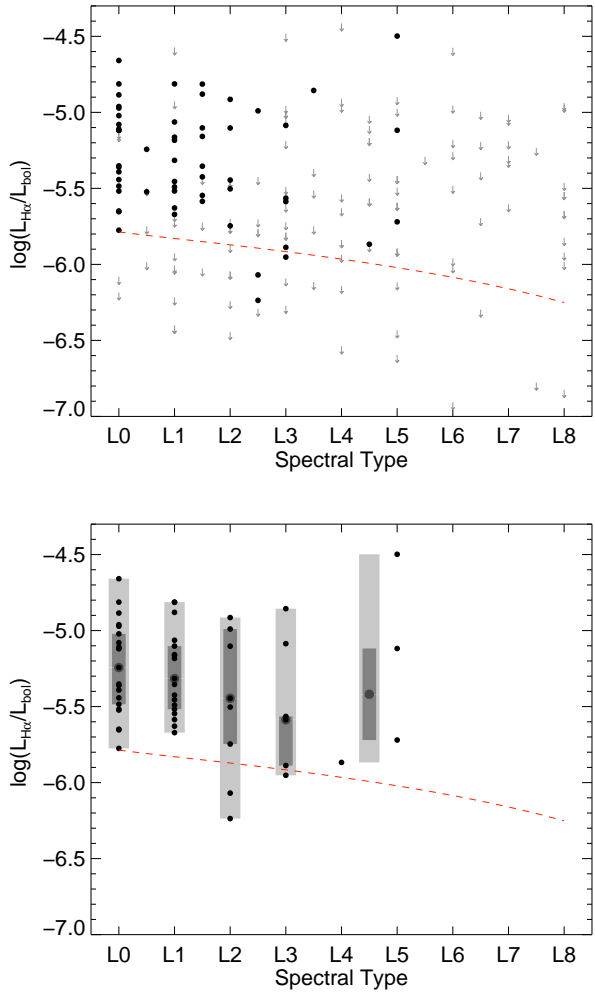} 
\caption[Activity strength as a function of spectral type.]{Activity strength of L dwarfs as a function of spectral type. Top panel: Activity strength detections (black circles) and upper limits (grey arrows), as reported in Table~\ref{tab:Lha}, are both shown. Bottom panel: Activity strength of H$\alpha$ detections only (smaller black circles), with half spectral types rounded down. For each spectral type bin (L4 and L5 are treated as a single bin) the full range of detections (light grey bar), the inter-quartile range (medium grey bar) and the median (larger dark grey circle) are shown, corresponding to the values shown in Figure~\ref{fig:lha_all} and used in Section~\ref{sec:chr_results}. In both panels, the values corresponding to the SDSS H$\alpha$ detection limit of EW$ \geq 0.75$~\AA~are shown (red dashed line).} \label{fig:lha_L}
\end{figure}

For each L dwarf, we recalculated the log($L_{\rm H\alpha}/L_{\rm bol}$) based on the published EW and the $\chi$ factors from \citet{Schmidt2014a}, which are higher than previous calculations of the $\chi$ factor \citep[e.g.,][]{Reiners2008}. The activity strengths of some L dwarfs are therefore larger than previously reported values. The resulting log($L_{\rm H\alpha}/L_{\rm bol}$) values are given in Table~\ref{tab:Lha} and shown in Figure~\ref{fig:lha_L}. The median activity strength is also shown compared to the M dwarf activity in Figure~\ref{fig:lha_all}. The median activity strength of L dwarfs decreases from log($L_{\rm H\alpha}/L_{\rm bol})=-5$ (L0) to $-5.7$ (L3). 

The L5 dwarfs show a higher level of activity compared to the earlier-L dwarfs. 2MASS 01443536$-$0716142 was observed to be active only during a short series of observations classified as a flare \citep[the value shown in Figure~\ref{fig:lha_L} is the minimum measured value from that flare;][]{Liebert2003}, and the high level of activity in 2MASS J1315$-$2649 has been discussed by \citet{Gizis2002b}, \citet{Hall2002} and \citet{Burgasser2011b}. LHS 102B shows weaker emission than the other two L5 dwarfs, and is closer to the activity strength that would be expected if the decline with spectral type extends through mid-L dwarfs. 

The L dwarf activity sample contains a heterogeneous mix of data from different surveys with varying S/N. To obtain a rough fraction of active dwarfs for this sample, we adapted the criteria described in Section~\ref{sec:habud} to use only the given detections and upper limits. We defined active dwarfs as those with H$\alpha$ EW $>0.75$~\AA~and inactive dwarfs as those with upper limits (non-detections) of H$\alpha$ EW $\leq0.75$~\AA. This approach ignores high upper limits (which do not place strong constraints on the H$\alpha$ emission) and weak detections (to allow a relatively uniform lower limit). The L dwarf activity fraction calculated using these criteria is shown in Figure~\ref{fig:frac}. The L dwarf activity sample does include data from the BUD L dwarfs, which make up 21\% of the total sample.

The L dwarf activity sample fractions agree with those solely based on the BUD L dwarfs, showing a decline from $\sim$90\% at L0 to $\sim$60\% at L3. The activity fractions for the L4 (33\%) and L5 (50\%) dwarfs are smaller, indicating the fraction of L dwarfs with H$\alpha$ emission continues to decrease. It is remarkable, however, that of the 84 L0-L5 dwarfs included in the activity fraction calculation, 54 (64\%) are active. Previous results indicated a sharp decline from late-M dwarfs to an absence of activity in mid-L dwarfs \citep[e.g.,][]{Gizis2000,Schmidt2007,Reiners2008}, but that result was likely an effect of smaller numbers or low S/N in the region surrounding H$\alpha$. While no activity is currently observed on L6-L8 dwarfs, that may be due to the small number of late-L dwarf spectra with sufficient S/N to detect H$\alpha$ emission. 

\section{The Temperature Structure of a Quiet Chromosphere}
\label{sec:tstr}
To characterize the chromospheres that give rise to the H$\alpha$ emission described in Section~\ref{sec:haobs}, we adopted the approach that has been used previously to investigate the chromospheres of M dwarfs \citep[e.g.,][]{Hawley2003,Fuhrmeister2005,Walkowicz2009}, using a one-dimensional approximation of the chromosphere (temperature as a function of column mass) to generate the H$\alpha$ emission. That H$\alpha$ emission in linear combination with the photospheric emission produces the observed spectrum of an active star or brown dwarf. 

The use of one-dimensional stellar atmospheres includes two main assumptions: 1) the temperature structure at each point on the surface of the star is either low (photospheric) or high (chromospheric) and 2) the chromosphere is well modeled by a single temperature distribution as a function of column mass. The first assumption has a basis in stellar bifurcation \citep{Ayres1981}; any material not heated to chromospheric temperatures will quickly cool to photospheric temperatures. The second assumption is an over-simplification, but our H$\alpha$ data alone cannot place meaningful constraints on lateral inhomogeneity, i.e., a chromosphere that changes temperature distribution as a function of atmospheric height over its spatial extent. Multiple emission lines are needed to characterize a multi-component chromosphere \citep[e.g.,][]{Walkowicz2008b}. 

\subsection{Constructing Model Atmospheres}
\label{sec:test}
We adopt the photospheric temperature structure of the BT-Settl models \citep{Allard2011}, which were generated with the LTE radiative transfer code Phoenix \citep{Hauschildt1999}. These models include a treatment of mixing and convection due to the sedimentation of the dust clouds that are common in L dwarf atmospheres. Additionally, the BT-Settl grid is continuous over the transition from M to L dwarf atmospheres (and beyond; the grid ranges from T$_{\rm eff} = 100,000$~K to T$_{\rm eff}= 400$~K). While the models also include $5.5>$log(g)$>-$0.5 and $0.5>$[M/H]$>-$1.5, for our initial investigation we assume a single log(g) = 5.0 and [M/H] = 0.0. To explore a range of M and L dwarfs, we select photospheres with five different representative effective temperatures, T$_{\rm eff}$ = 1400, 1900, 2400, 2900, and 3400~K, roughly corresponding to L7, L3, M8, M4, and M0 dwarfs respectively \citep{Stephens2009,Rajpurohit2013}. 

\begin{figure}[t]
\includegraphics[width=\linewidth]{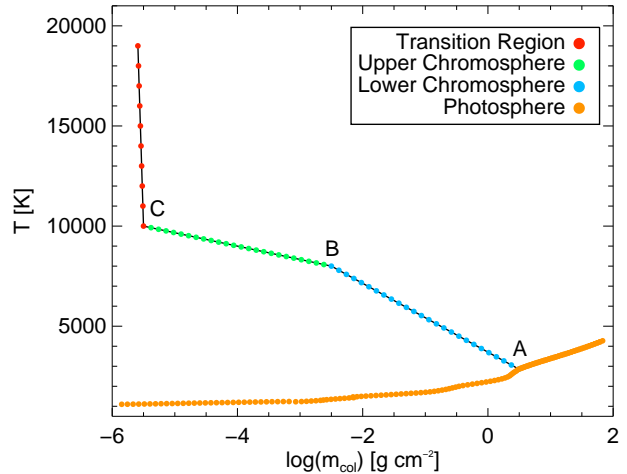} 
\caption[Model atmosphere of a T$_{\rm eff}$ = 2400K dwarf with an added chromosphere.]{Photospheric model atmosphere of T$_{\rm eff}$ = 2400~K with an added two component chromospheric temperature structure. The different regions of the atmosphere are shown as colored circles (see legend). The features that are discussed throughout Section~\ref{sec:test} are labeled: the base of the chromosphere (A), the chromosphere break (B), and the beginning of the transition region (C).} \label{fig:oneatmo}
\end{figure}

For each model, we replaced the temperature in the outer atmosphere with a chromospheric temperature structure consisting of two components, each modeled by a linear increase in temperature with the log of column mass (log(m$_{\rm col}$)). Figure~\ref{fig:oneatmo} illustrates these components, which are characterized by the positions of the start of the chromospheric temperature rise (A), the chromosphere break\footnote{In this paper, the chromosphere break is used as a convenient parameter to characterize the chromospheric temperature structure. The motivation for the two different temperature slopes with respect to column mass can be found in detailed models of the Sun's chromosphere \citep[e.g.][]{Vernazza1981}. The lower chromosphere cannot cool efficiently so the temperature rises steeply as a function of column mass, while the upper chromosphere cools via hydrogen emission resulting in a shallower temperature structure.} (B), and the start of the transition region (C). This chromospheric temperature structure is based on previous model chromospheres of M dwarfs \citep[e.g.,][]{Fuhrmeister2005,Walkowicz2008b} which are analogs of solar chromosphere models \citep[e.g.,][]{Vernazza1981}.

We produced a grid of model atmospheres by changing the locations (A, B, C) in log(m$_{\rm col}$) and temperature. The location of the base of the chromosphere (A in Figure~\ref{fig:oneatmo}) has no effect on the H$\alpha$ flux; the material is too cool to emit or absorb at the H$\alpha$ transition except in the hottest T$_{\rm eff}$ models. In those hot models, the added H$\alpha$ emission is small compared to the observed range of H$\alpha$ emission. Because changing the base of the chromosphere has little effect on the H$\alpha$ emission, we fixed the base of the chromosphere at log(m$_{\rm col}$) = 0.5 g cm$^{-2}$ for all models, a value where there is little contribution to H$\alpha$ even for the hotter T$_{\rm eff}$ models\footnote{The temperature at the start of the chromosphere is given by the temperature of the photosphere model at this log(m$_{\rm col}$).}. 

The position of the start of the transition region (C in Figure~\ref{fig:oneatmo}) in both temperature and log(m$_{\rm col}$) has little effect on the H$\alpha$ flux. That material is too diffuse to produce a significant amount of H$\alpha$ emission compared to material deeper in the atmosphere. We fixed the beginning of the transition region at T$=10,000$~K and log(m$_{\rm col}$) = $-5.5$ g cm$^{-2}$ and included an increase to T$=19,000$~K between log(m$_{\rm col}$) = $-5.5$ and $-5.6$ g cm$^{-2}$ to mimic the beginning of the transition region \citep[following the prescriptions of][for earlier type M dwarfs]{Fuhrmeister2005,Walkowicz2008b}.

\begin{figure*}
\includegraphics[width=\linewidth]{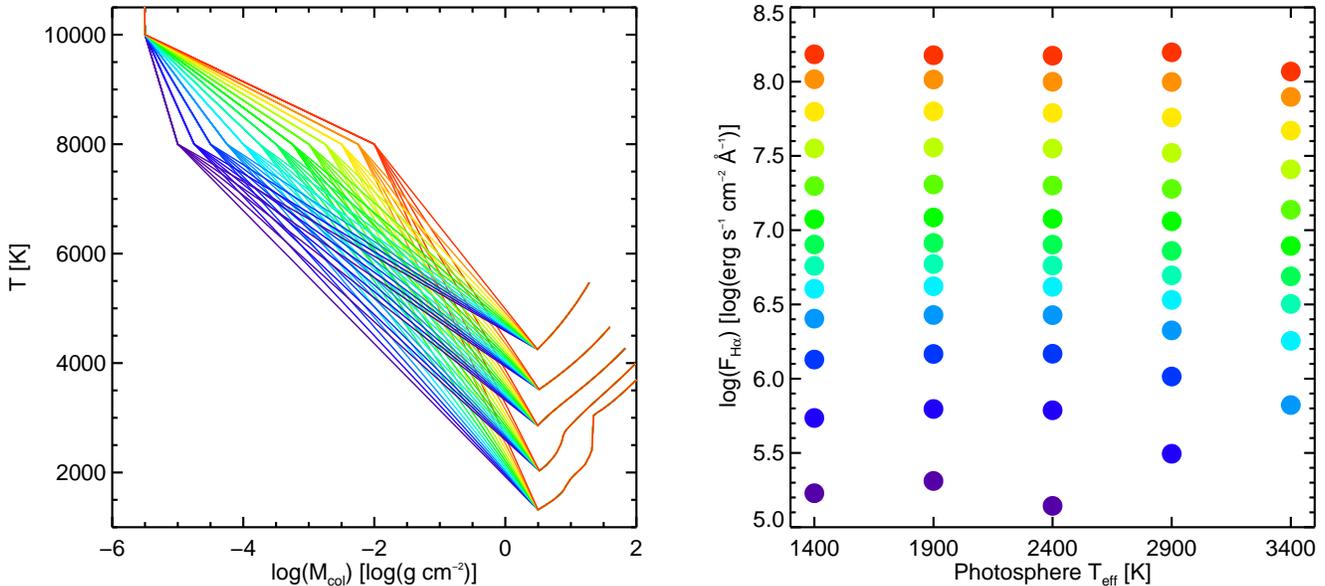} 
\caption[A grid of model atmospheres and their resulting H$\alpha$ flux.]{Left panel: The grid of model atmospheres used to generate H$\alpha$ emission. The atmosphere grid is based on model photospheres with five different T$_{\rm eff}$ (3400, 2900, 2400, 1900, and 1400~K from top to bottom, roughly corresponding to M0, M4, M8, L3, and L7 dwarfs, respectively) with thirteen different chromospheric temperature structures attached at log(m$_{\rm col}$)=0.5. Each chromosphere break is shown in a different color; the hottest chromospheres are shown in red and the coolest chromospheres are in purple. Right panel: The log of H$\alpha$ surface flux as a function of photosphere T$_{\rm eff}$, with results from each chromosphere break represented by the same color as in the left panel. H$\alpha$ lines that were in absorption are not shown.} \label{fig:atflux}
\end{figure*}

The position of the chromosphere break (B in Figure~\ref{fig:oneatmo}) has a large and direct effect on the H$\alpha$ flux. Material at T$\sim8000$~K produces significant emission, due primarily to the high fractional ionization of hydrogen at T$\sim8000$~K \citep[e.g.,][]{Cram1979}. Varying the log(m$_{\rm col}$) at which the chromosphere reaches T$=8000$~K results in models with a large range of H$\alpha$ flux to compare with the data. We produced our grid of chromospheres by varying the location in log(m$_{\rm col}$) of the chromosphere break from $-5$ to $-2$ g cm$^{-2}$. Moving the chromosphere break deeper in the atmosphere (towards log(m$_{\rm col}$)$\sim-2$~g~cm$^{-2}$) creates a \textit{hotter} chromosphere due to the higher density of hot (T$\sim8000$~K) material; \textit{cooler} chromospheres (chromosphere breaks closer to log(m$_{\rm col}$)$\sim-5$~g~cm$^{-2}$) have a lower density of material heated to T$=8000$~K. The grid of these 13 different chromospheres attached to five different underlying photospheres is shown in the left panel of Figure~\ref{fig:atflux}.

\subsection{Using Chromospheric Models to Calculate Activity Strength}
\label{sec:RH}
The output H$\alpha$ flux for each atmosphere model was calculated using the RH radiative transfer code \citep{Uitenbroek2001}. RH is based on the Multilevel Accelerated Lambda Iteration (MALI) formalism of \citet{Rybicki1991}. Lambda Iteration is a method of treating non-LTE (local thermal equilibrium) effects by iteratively calculating the radiation field (based on local populations) and the local populations (which in turn depend upon the radiation field). This process is repeated until the relative change between iterations is less than 0.01 for both the populations and output radiation field.

To calculate the radiation field, RH relies upon the formalism of Partial Redistribution (PRD), which allows the optically thick cores of lines to be in LTE, but treats the NLTE effects of coherent scattering in the optically thin line wings. PRD has been demonstrated to greatly speed the convergence of MALI on a final solution \citep[e.g.][]{Paletou1995}. During the calculations, one atom is solved in detail, while the rest are treated as background opacity. RH has been implemented with one-, two- and three- dimensional geometries, with the complexities of multidimensional modeling made possible by the speed of PRD iterations, but is limited to considering a static system rather than computing the full magneto-hydrodynamics of the chromosphere. 

To model the H$\alpha$ flux for cool and ultracool dwarfs, we used one-dimensional geometry and a six-level hydrogen atom. As discussed above, limiting the calculations to one dimension simplifies the likely complex structure of the chromosphere over the stellar surface. The model is also static and does not include the heating mechanisms that shape the chromosphere temperature structure as a function of height. 

For each model, the RH output spectrum provides the flux at the stellar surface. We measured the H$\alpha$ surface flux from the output spectrum by integrating the flux (above the continuum) from 6540~\AA~to 6586~\AA. This relatively large wavelength range was required by the hottest model chromospheres, where the resulting emission lines were quite broad. The resulting H$\alpha$ surface fluxes are shown in the right panel of Figure~\ref{fig:atflux}. The strength of the line flux over the range of T$_{\rm eff}$ depends primarily on the position of the chromosphere break, so the H$\alpha$ surface flux is relatively constant as a function of photosphere T$_{\rm eff}$. The exceptions to constant H$\alpha$ surface flux are due to absorption in the lower chromosphere. 

For early- to mid-M dwarfs, cooler chromospheres result in H$\alpha$ absorption rather than emission. The absorption originates in the lower chromosphere, where it is hot enough to populate the $n=2$ level of hydrogen, but the level population is not collisionally dominated \citep[as it is near the chromosphere break,][]{Cram1979,Robinson1990}. In our models, the coolest two chromospheres for the T$_{\rm eff}=3400$~K photosphere and the coolest chromosphere on the T$_{\rm eff}=2900$~K photosphere show absorption, not emission. This absorption also decreases the total H$\alpha$ surface flux from the hottest chromospheres, causing the slight decline shown in Figure~\ref{fig:atflux}. The cooler photosphere models also have cooler lower chromospheres so do not populate enough of the $n=2$ level for absorption. 

\begin{figure*}
\includegraphics[width=0.95\linewidth]{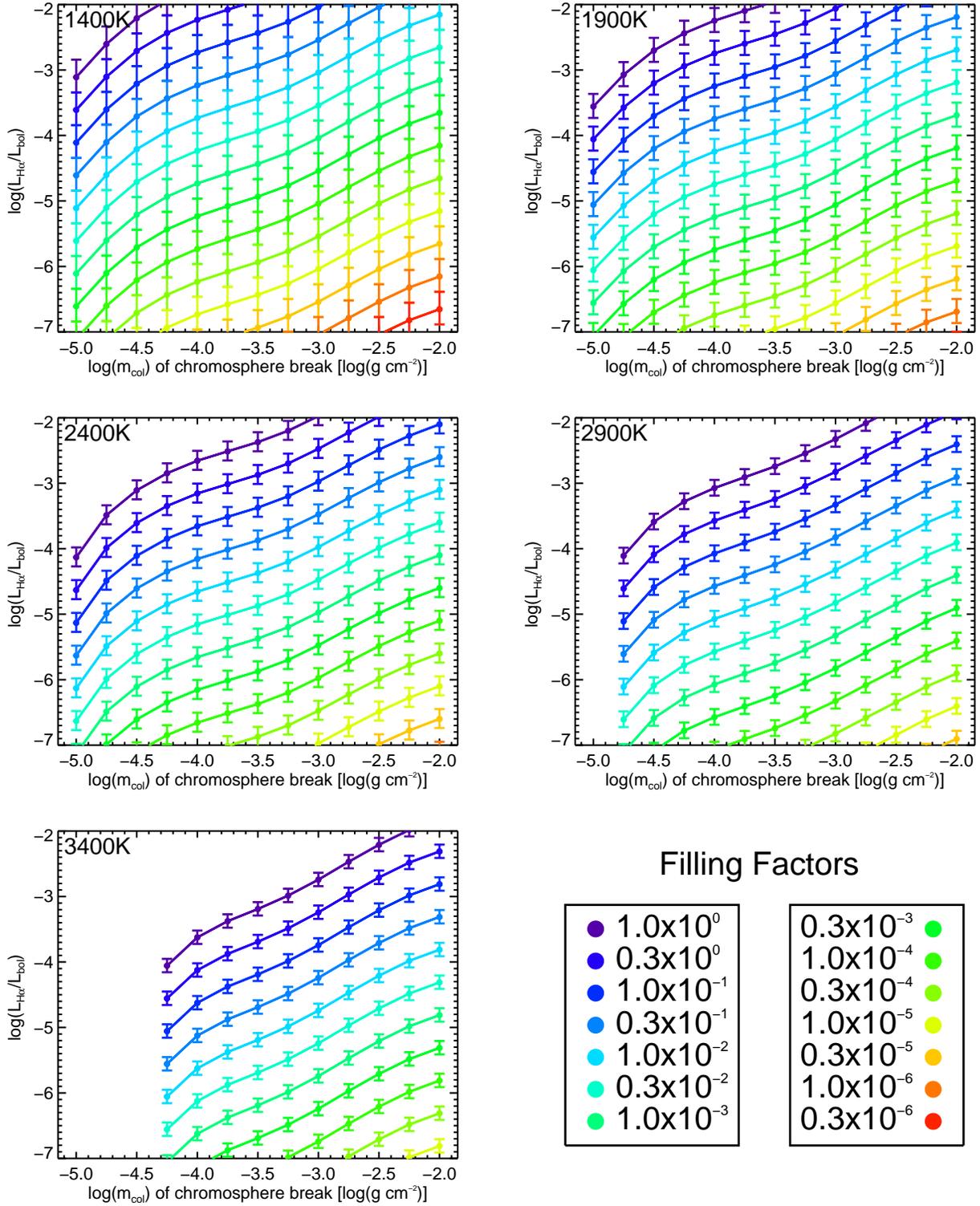} 
\caption[Activity strength as a function of chromosphere break for a range of chromospheric filling factors for each photosphere temperature.]{For each photosphere T$_{\rm eff}$, log($L_{\rm H\alpha}/L_{\rm bol}$) is shown as a function of chromosphere break for a range of chromospheric filling factors (colors given in the bottom right). The uncertainties in log($L_{\rm H\alpha}/L_{\rm bol}$) are based on varying T$_{\rm eff}$ by $\pm$200K, as discussed in Appendix~A.} \label{fig:lum}
\end{figure*}

The model flux can be compared to the observed H$\alpha$ emission if we estimate the chromospheric filling factor (surface coverage). We used a range of filling factors and estimated the bolometric fluxes based on the model T$_{\rm eff}$ (as discussed in Appendix~A) to obtain log($L_{\rm H\alpha}/L_{\rm bol}$). The results are shown in Figure~\ref{fig:lum} for each T$_{\rm eff}$, chromosphere break, and a range of chromospheric filling factors from 3$\times10^{-7}$ to 1.0. The H$\alpha$ emission strength declines with cooler chromospheric temperatures and smaller filling factors. Additionally, the strength of emission produced at the same chromosphere break and chromospheric filling factor declines for hotter photospheric temperatures. For example, the hottest chromosphere with a chromospheric filling factor of 0.01 produces log($L_{\rm H\alpha/}L_{\rm bol})\sim-2$ for T$_{\rm eff}$ = 1400K and log($L_{\rm H\alpha}/L_{\rm bol})\sim-4$ for T$_{\rm eff}$ = 3400K. As the line flux remains nearly constant for the different values of T$_{\rm eff}$ (see Figure~\ref{fig:atflux}) in our models, this effect is primarily due to the decrease in bolometric luminosity for the cooler objects. The ratio between the line luminosity and the total luminosity calculated based on the grid of models increases rather sharply for these cool objects because their total luminosity decreases, not because of the increased line flux.

\section{Chromospheres of M and L dwarfs}
\label{sec:chr_results}
\begin{figure}
\includegraphics[width=\linewidth]{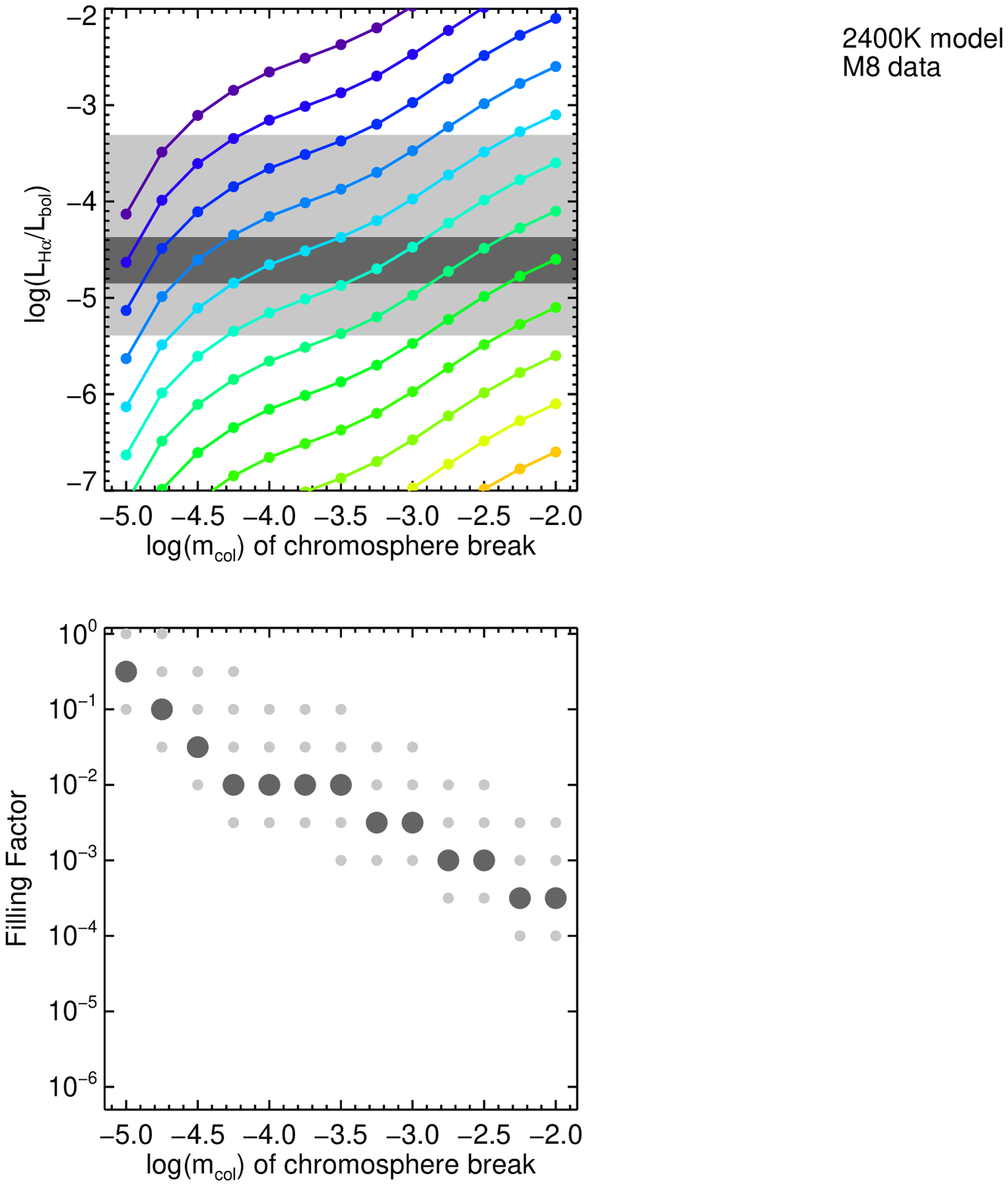} 
\caption[A comparison of generated activity strength with observations and the constraints it places on chromospheric filling factor and temperature structure.]{Top panel: log($L_{\rm H\alpha}/L_{\rm bol}$) as a function of log(m$_{\rm col}$) of the chromosphere break for the T$_{\rm eff} =2400$~K set of models. The colored lines show the same chromospheric filling factors as in Figure~\ref{fig:lum}. The light grey shading represents the full range of observed log($L_{\rm H\alpha}/L_{\rm bol}$) for M8 dwarfs, and the dark grey shading indicates the interquartile range. Bottom panel: Filling factor as a function of log(m$_{\rm col}$) of the chromosphere break. Values that correspond to models that fall in the full range of observations for M8 dwarfs are indicated by small light grey circles, and values in the interquartile range are the large dark grey circles.} \label{fig:demo}
\end{figure}

In the previous two sections, we characterized the activity strength of M and L dwarfs based on both observations and a suite of chromosphere models. Figure~\ref{fig:demo} displays the comparison of log($L_{\rm H\alpha}/L_{\rm bol}$) generated from the full range of chromospheres on the T$_{\rm eff}=2400$~K model and the M8 data. We show the ranges of chromospheric filling factor and log(m$_{\rm col}$) of the chromosphere break where the model and data overlap. The allowed filling factors for the observed range of H$\alpha$ emission for M8 dwarfs range from 10$^{-4}$ to 1, but the specific value of filling factor depends on chromosphere temperature structure. Because the cooler chromospheres have less H$\alpha$ surface flux, they must cover a larger fraction of the surface to produce the same total activity strength as the hotter chromospheres. Without independent constraints on the chromosphere temperature structure or the filling factor, we can only examine the range of both. 

\begin{figure}
\includegraphics[width=\linewidth]{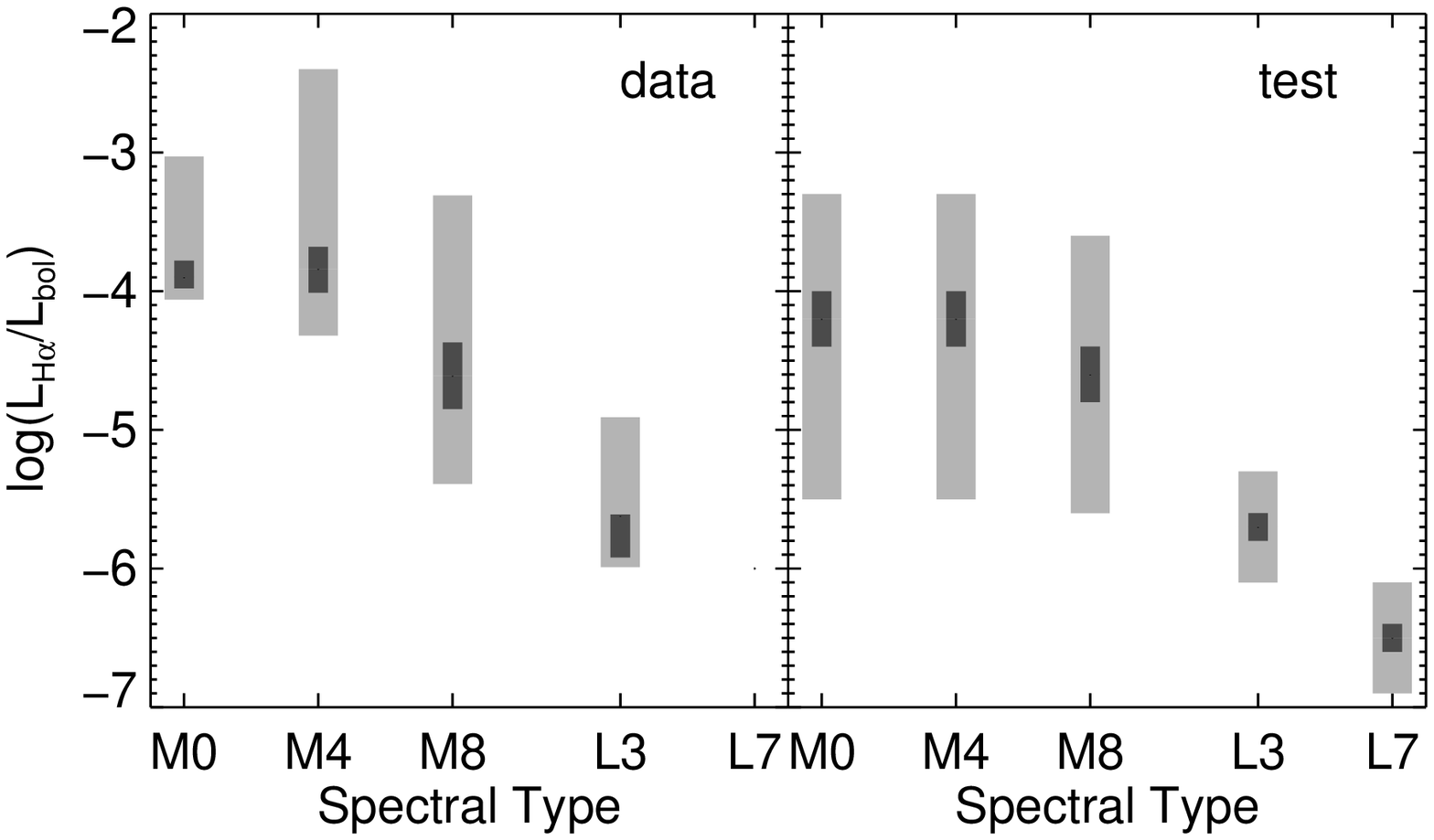} 
\caption[Data and smoothed test values for activity strength as a function of spectral type.]{Values of log($L_{\rm H\alpha}/L_{\rm bol}$) as a function of spectral type. The inner (dark grey) shaded box represents the interquartile range, while the outer (light grey) shaded box represents the entire range. Left panel: values taken from Figure~\ref{fig:lha_all}, selected to overlap with the T$_{\rm eff}$ values of the model grid. Right panel: values that have been modified to examine the effect of the biases described in Section~\ref{sec:chr_results}.} \label{fig:valrng}
\end{figure}

The range of filling factors and chromosphere temperature structures consistent with the data changes over the entire M and L spectral classes. To examine the chromospheres for M and L dwarfs, we used the same comparison of models and data as described for the M8 dwarfs above. To account for the biases described in Section~\ref{sec:habud}, we adopted two slightly different sets of data-based comparison values, shown in Figure~\ref{fig:valrng}. The first set of values (``data") are taken directly from the log($L_{H\alpha}/L_{\rm bol}$) as a function of spectral type shown in Figure~\ref{fig:lha_all}. The second set of values (``test") are based on that data but were modified to examine the effects of observational biases (discussed further in Sections~\ref{sec:chrM} and~\ref{sec:chrL}). For the ``test" values, the minimum activity strengths for M0 and M4 were reduced to include possible emission below the H$\alpha$ EW$\geq$0.75~\AA~limit, while the maximum value of the M4 bin was reduced as the strongest H$\alpha$ emission could be due to flares rather than quiescent emission. An estimate of L7 activity strength was also added to the ``test" values, selected based on the observed upper limits for emission. 

\begin{figure*}
\includegraphics[width=0.475\linewidth]{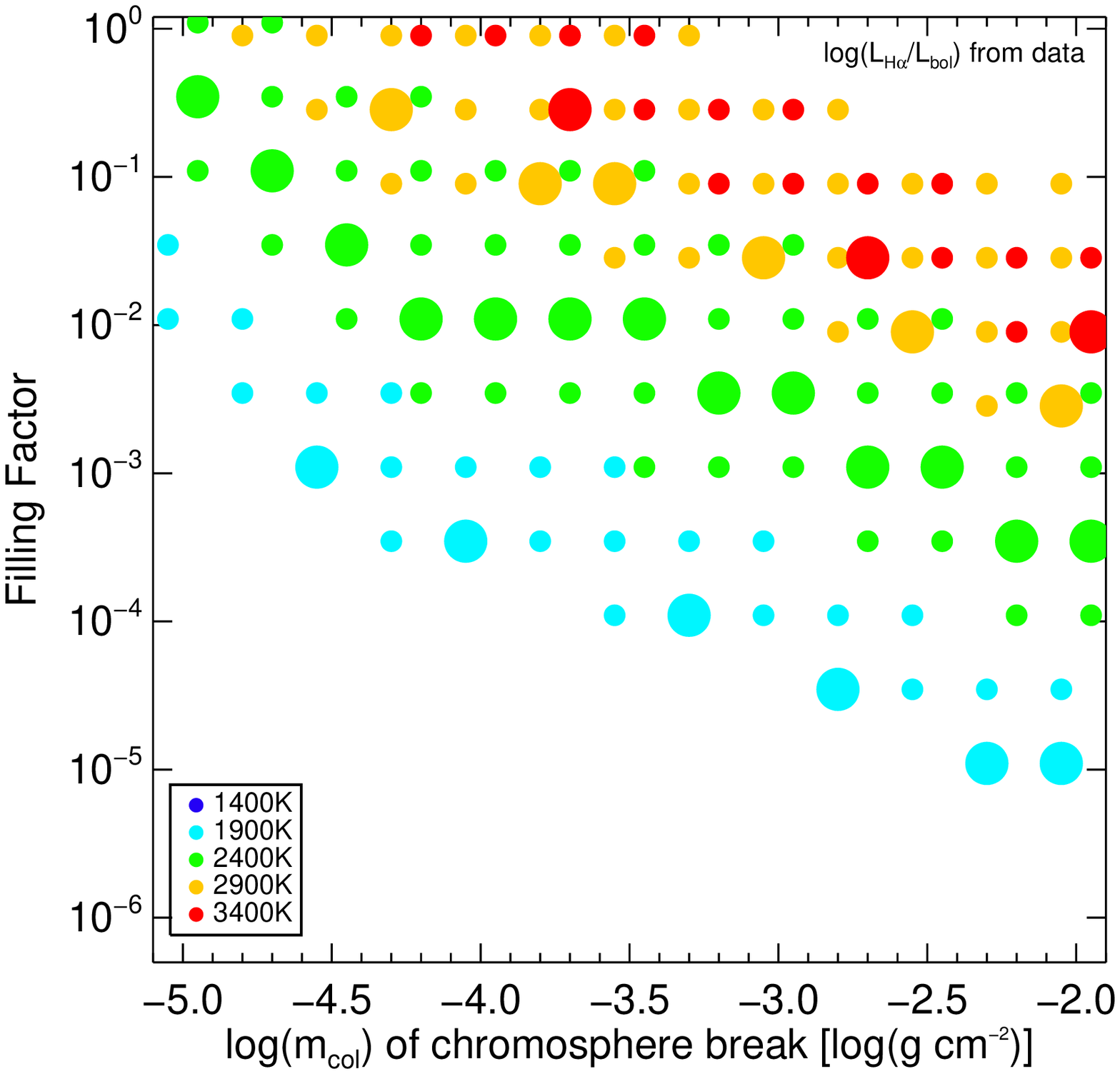} 
\includegraphics[width=0.475\linewidth]{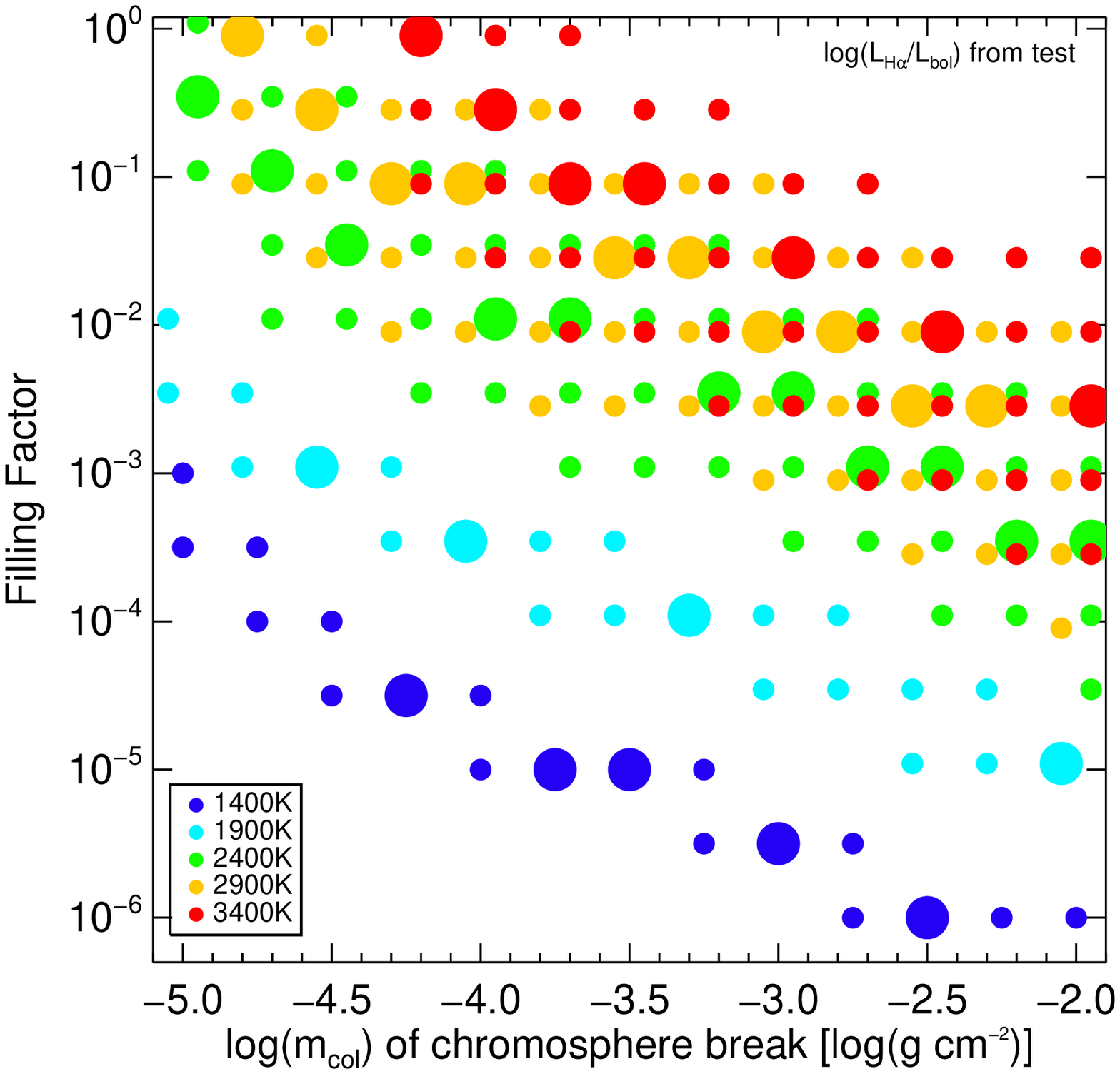} 
\caption[Chromosphere break as a function of chromospheric filling factor for models that match observations.]{Chromosphere break as a function of chromospheric filling factor for models that match the ranges of log($L_{\rm H\alpha}/L_{\rm bol}$) for each T$_{\rm eff}$/Spectral Type shown in Figure~\ref{fig:valrng}. For each T$_{\rm eff}$, the large circles are values that match the inner quartile range of observed values, while the smaller circles show matches with the entire range of values. Circles are slightly offset from their exact values for clarity. The left panel corresponds to the left panel of Figure~\ref{fig:valrng}, illustrating the models that best match the values adopted from the H$\alpha$ emission data. The right panel corresponds to the right panel of Figure~\ref{fig:valrng}, showing the models that best match the ``test'' values adopted to examine the effect of the biases (e.g., inclusion of serendipitous flare emission from M4 dwarfs and the H$\alpha$ detection limit in SDSS) described further in Section~\ref{sec:chr_results}.} \label{fig:results}
\end{figure*}

Figure~\ref{fig:results} presents the results of comparing the ``data" and ``test" values with the models shown in Figure~\ref{fig:lum}. As for the M8 data and the T$_{\rm eff}=2400$~K models (shown in Figure~\ref{fig:demo}), we marked the combinations of chromosphere break and filling factor consistent with the full range of data as well as the combinations that match with the inter-quartile range. For each spectral type/T$_{\rm eff}$, there is a spread of 3--4 orders of magnitude in filling factor over the full range of chromosphere models examined. The only chromosphere models entirely inconsistent with the observed values are the coolest models on the M0/T$_{\rm eff}=3400$~K photosphere. 

For each single model (fixing both T$_{\rm eff}$ and chromosphere break), a filling factor variation of 1--2 orders of magnitude will produce the observed range of activity range at each spectral type. At fixed filling factor, the same activity range covers many different chromosphere breaks. There is a trend with spectral type/T$_{\rm eff}$: the early-M dwarf chromospheres are generally more extended and/or hotter than those of late-M dwarfs, which also have more extended and/or hotter chromospheres than L dwarfs. The differences between the ``test" and ``data" values are small. In the ``data" results, the M0 and M4 overlap more in filling factor and chromosphere break, while the M8 dwarfs were produced by cooler or less extended chromospheres. In the ``test" results, the M0 and M4 overlap, instead showing a gradual change over the M spectral type. In the next three subsections, we discuss these results in detail.

\subsection{The Chromospheres of M dwarfs}
\label{sec:chrM}
When considering H$\alpha$ emission across the M spectral subclasses, it is important to consider both the fraction of active M dwarfs and the strength of observed emission. As shown in Figure~\ref{fig:frac}, the fraction of M0 and M4 dwarfs that posses H$\alpha$ emission (with EW$>0.75$~\AA) is 2\% and 12\%, respectively, compared to 80\% for M8 dwarfs. The fractions of active M0-M4 dwarfs are likely to be higher than these observations indicate due primarily to the exclusion of H$\alpha$ absorption as a chromosphere tracer in this dataset. As discussed in Section~\ref{sec:RH}, a cooler chromosphere on an early-M dwarf will produce absorption, while cooler chromospheres on late-M and L dwarfs show weak emission. We do not present the models consistent with H$\alpha$ absorption because activity strength cannot be quantified with the $L_{\rm H\alpha}/L_{\rm bol}$ metric for absorption.

The detection of M0 and M4 dwarfs with H$\alpha$ absorption provides an interesting constraint on the chromosphere breaks of M0 and M4 dwarfs. H$\alpha$ absorption is produced by models of M0 dwarfs with a chromosphere break of log(m$_{\rm col}$) $<-4$~g~cm$^{-2}$ and M4 dwarfs with a chromosphere break of log(m$_{\rm col}$) $<-4.5$~g~cm$^{-2}$. Thus, some fraction of early- to mid-M dwarfs \textit{must} have cooler chromospheres to produce the observed H$\alpha$ absorption. These stars are excluded from the SDSS H$\alpha$ emission data, while ultracool dwarfs with similarly cool chromospheres would be included. More generally, this result also indicates that there is some variation in the chromosphere temperature structure; variations between emission strength within a spectral type are not due only to differences in filling factor. 

The comparison of the data and test values (shown in Figure~\ref{fig:valrng}) to the models (Figure~\ref{fig:results}) show the changes in M dwarf chromospheres with spectral type. The data values of chromosphere break and filling factor overlap for M0 and M4 dwarfs, in part due to the large range of M4 H$\alpha$ emission. The M8 filling factors are typically an order of magnitude smaller than for the earlier M dwarfs. When the M0 and M4 values are adjusted downward in the ``test" set (to include weak H$\alpha$ emission and exclude flares), the M dwarf chromospheres have similar ranges of chromosphere break and filling factor and display a smooth progression with spectral type. Initial measurements of late-M dwarf magnetic fields show that they are similar in strength to those of early- to mid-M dwarfs \citep[e.g.,][]{Reiners2007a}. The model chromosphere results suggest that the heating of late-M dwarf chromospheres remains strong and is comparable to the earlier M dwarfs. 

\subsection{The Chromospheres of L dwarfs}
\label{sec:chrL}
The fraction of L3 dwarfs with H$\alpha$ emission $>0.75$~\AA~is relatively unconstrained, but likely to be over 50\% (Figure~\ref{fig:frac}). The median activity strength falls close to the 0.75~\AA~limit, so it is possible that the fraction is an underestimate of the total number of mid-L dwarfs with chromospheres (excluding weakly active L3 dwarfs, see Figure~\ref{fig:lha_L}). As shown in Figure~\ref{fig:results}, the chromospheres of the observed L3 dwarfs cover significantly less of the surface (at the same chromosphere break, 10 times smaller) or are much cooler (at similar filling factors, a change of $-2$ in the log($m_{col}$) of the chromosphere break) than M8 dwarfs, indicating a shift in the chromospheric structure between late-M and early-L dwarfs. The shift is slightly larger for the test values, where the L3 range was adjusted to include possible weaker emission falling beneath the detection limit. 

There are no direct measurements of the magnetic fields generated in L3 dwarfs, but the few early-L dwarfs with detectable X-ray \citep{Audard2007} and radio \citep{Hallinan2008,Williams2014} emission suggest their surface magnetic fields are similar to those found on late-M dwarfs. If the magnetic fields are similar, then the significant difference in the average chromosphere temperature structure and/or filling factor between M8 and L3 dwarfs may be due to less efficient chromospheric heating. 

Beyond spectral type L3, there are only four detections of H$\alpha$ emission from L dwarfs. Many L dwarfs have optical spectra covering that region; there are 10 upper limits at or below the SDSS cutoff of 0.75~\AA~and 45 higher upper limits (due to the faintness of the continuum in that region, it is difficult to obtain sufficient S/N). There are a handful of upper limits sufficiently low to indicate the absence of a chromosphere \citep[e.g., $<$0.2~\AA~upper limits from][]{Reiners2008}, but if the activity strength continues its decline with respect to spectral type, the median activity level for late-L dwarfs would fall at or below the 0.75~\AA~limit. We examined models consistent with a low activity strength level for L7 dwarfs, as shown in the test values of Figure~\ref{fig:valrng}. If L7 dwarfs do have H$\alpha$ emission, it is generated in a chromosphere covering a very small fraction of the surface (filling factors of $<10^{-4}$). 

\subsection{Discussion}
\label{sec:disc}
The total picture of activity on M and L dwarfs is complex. Each spectral type bin from M0 to L5 includes a wide range of observed activity strength (including both those without detected H$\alpha$ emission and measurements spanning one to two orders of magnitude), while L6--L8 dwarfs have either very weak or no activity. The fraction of dwarfs with observed H$\alpha$ emission increases from 2\% at M0 to 90\% at L0, then declines to 50\% at L5. Activity strength, which is constant for M0-M4 dwarfs at log($L_{\rm H\alpha}/L_{\rm bol}$)$ \sim -3.8$, declines to log($L_{\rm H\alpha}/L_{\rm bol}$)$ = -5.7$ at spectral type L3. The range of chromospheres that produce the observed activity strength is similar for M0--M8 dwarfs, but the chromospheres consistent with the H$\alpha$ observed from L3 dwarfs are much cooler and/or less extended. Based on the results of \citet{West2008}, we know that the increase in activity fraction through the M spectral type range is due to the changing relationship between mass, age, and activity. The reason for the decline in activity for L dwarfs is instead due to the decreasing ionization in the cool photospheres. While extremely weak activity on L6--L8 dwarfs is indistinguishable from no activity at the resolution of our spectroscopic sample, the two scenarios represent distinct physical situations. 

\citet{Mohanty2002} described the interaction between surface magnetic fields and chromospheres on ultracool dwarfs using the magnetic Reynolds number ($R_m$), which quantifies the resistivity of the gas to interaction with the magnetic field. A $R_m<$1 indicates there is too much resistivity for magnetic fields to be important in fluid motions, so magnetic heating in a plasma with $R_m<$1 is depressed. $R_m$ changes with atmospheric height, and \citet{Mohanty2002} calculated the average $R_m$ for the atmospheres of T$_{\rm eff}=1500$~K to $3000$~K throughout the atmosphere. $R_m$ numbers are well below one at the surface for atmospheres of T$_{\rm eff}<2300$~K (spectral types M9/L0 and later), and $R_m$ is only larger than one deep into or below the photosphere ($\tau>$ 2 or \mbox{log(col. mass) $>$ 0.2}). In Figures~\ref{fig:frac} and~\ref{fig:lha_all}, we show this T$_{\rm eff}<2300$~K threshold compared to the activity fraction and activity strength as a function of spectral type. 

When interpreted in the context of previous data \citep[which showed a decreasing activity fraction for M8 and later spectral types, e.g.,][]{Gizis2000}, the T$_{\rm eff}<2300$~K limit suggests that activity was not possible in L dwarfs. However, our data and chromosphere models discussed above indicate that the consequence of lower $R_m$ numbers in the photospheres of L dwarfs is a chromosphere with a smaller filling factor and cooler chromospheric break (see Section~\ref{sec:chrL}). If the $R_m$ calculations are correct for average L dwarf parameters, four likely possibilities for activity on L dwarfs include: 1) a localized increase in the magnetic field \citep[e.g., a dwarf with a non-axisymmetric field with significant spatial variation;][]{Morin2010}; 2) localized backwarming due to patchy clouds \citep[common at the L/T transition; e.g.,][]{Marley2010,Radigan2012}, which heats a small portion of the surface and temporarily increases ionization allowing more efficient magnetic heating; 3) the creation of buoyant flux tubes from interactions between the magnetic field and the hotter, more ionized plasma further under the surface \citep{Mohanty2002}; or 4) heating the chromosphere from auroral interactions in the upper atmosphere \citep[e.g.,][]{Hallinan2008,Berger2009}.

Additionally, the \citet{Mohanty2002} calculations assumed that the charged particles in cool atmospheres are produced only from atomic ionization and ignored the effects of dust. \citet{Helling2011a,Helling2011b,Helling2013} explore the interaction between ionized dust particles in brown dwarfs and planetary atmospheres. Dust can interact to form temporary ``streamers" which ionize the surrounding medium. While these ``streamers" likely dissipate on timescales of minutes to hours, it is possible that this mechanism is important in the generation of transient chromospheric features on L dwarfs. 

\section{Summary}
\label{sec:sum}
In this paper, we outlined the selection of the BUD sample, 11,820 M and L dwarfs drawn from the spectroscopic data taken as part of SDSS DR7 and BOSS. The majority of these ultracool dwarfs have photometric data from SDSS, 2MASS, and WISE, enabling a full investigation of their colors. As discussed in S10, $i-z$ and $z-J$ show clear relations with spectral type (T$_{\rm eff}$) for late-M and early-L dwarfs. 2MASS and WISE colors have a weaker dependence on spectral type for these late-M and early-L dwarfs. We also present an extension of the \citet{Davenport2014} SDSS--2MASS-WISE color locus as a function of $i-J$ color, from $i-J$ =2.8 to 4.9. 

Using the spectroscopic data from BOSS combined with previous observations of M and L dwarfs, we examined the H$\alpha$ emission line across the M/L spectral sequence. We find that the fraction of active dwarfs increases from early-M to early-L spectral types. There are not sufficient data to determine if the activity fraction shows a steady or sharp decline from early-L to late-L spectral types and decreases at later spectral types. Activity strength, characterized by log($L_{\rm H\alpha}/L_{\rm bol}$), begins to decline at M4 and continues to decrease until spectral type L3. After the L3 spectral subclass, the median activity strength may continue to decline below our detection limit, or activity may simply not be present in the majority of mid- and late-L dwarfs.

We also used one-dimensional chromosphere models to estimate the temperature structures and filling factors of M and L dwarf chromospheres. The temperature structures and filling factors consistent with M dwarf observations are similar across the M spectral class (M0-M8). Early-L dwarfs have significantly weaker chromospheres than M dwarfs, likely due to the low ionization fractions in their cool atmospheres. The upper limits placed on late-L dwarf H$\alpha$ emission are only consistent with very cool and spatially confined (small surface coverage) chromospheres. 

\acknowledgements
J.\ J.\ B.\ acknowledges the financial support of NSF grant AST-1151462.
A.\ A.\ W.\ acknowledges funding from NSF grants AST-1109273 and AST-1255568 and the support of the Research Corporation for Science Advancement's Cottrell Scholarship.

This publication makes use of data products from the Two Micron All Sky Survey, which is a joint project of the University of Massachusetts and the Infrared Processing and Analysis Center/California Institute of Technology, funded by the National Aeronautics and Space Administration and the National Science Foundation. This publication also makes use of data products from the Wide-field Infrared Survey Explorer, which is a joint project of the University of California, Los Angeles, and the Jet Propulsion Laboratory/California Institute of Technology, funded by the National Aeronautics and Space Administration.

Funding for SDSS-III has been provided by the Alfred P. Sloan Foundation, the Participating Institutions, the National Science Foundation, and the U.S. Department of Energy Office of Science. The SDSS-III web site is \url{http://www.sdss3.org/}.

SDSS-III is managed by the Astrophysical Research Consortium for the Participating Institutions of the SDSS-III Collaboration including the University of Arizona, the Brazilian Participation Group, Brookhaven National Laboratory, Carnegie Mellon University, University of Florida, the French Participation Group, the German Participation Group, Harvard University, the Instituto de Astrofisica de Canarias, the Michigan State/Notre Dame/JINA Participation Group, Johns Hopkins University, Lawrence Berkeley National Laboratory, Max Planck Institute for Astrophysics, Max Planck Institute for Extraterrestrial Physics, New Mexico State University, New York University, Ohio State University, Pennsylvania State University, University of Portsmouth, Princeton University, the Spanish Participation Group, University of Tokyo, University of Utah, Vanderbilt University, University of Virginia, University of Washington, and Yale University. 

\appendix
\section{Converting Modeled Quantities to Observed Values}
For comparison of the model output with observations, we chose to convert H$\alpha$ line flux to log($L_{\rm H\alpha}/L_{\rm bol}$). As the ratio of two luminosities, log($L_{\rm H\alpha}/L_{\rm bol}$) has the advantage of not being dependent on the radius of each star or brown dwarf \citep[radii are poorly known for M and L dwarfs, especially over a range of magnetic activity, e.g.,][]{Kraus2011}.

$L_{\rm H\alpha}/L_{\rm bol}$ can be calculated from model parameters using:
\begin{equation}
\frac{L_{\rm H\alpha}}{L_{\rm bol}} = \frac{F_{\rm H\alpha} \times f \times 4 \pi R^2}{L_{\rm bol}},
\end{equation}
where $L_{\rm H\alpha}$ is the luminosity in the H$\alpha$ line, $L_{\rm bol}$ is the bolometric luminosity, $F_{H\alpha}$ is the flux in the H$\alpha$ line, $f$ is the chromospheric filling factor and $R$ is the stellar radius. 

We can estimate the bolometric luminosity based on the thermal emission for the star or brown dwarf's effective temperature and radius: 
\begin{equation}
L_{\rm bol} \approx L_{\rm thermal}= 4 \pi R^2 \sigma T_{\rm eff}^4 ,
\end{equation}
where $\sigma$ is the Stephan-Boltzmann constant. This approach includes the assumption that non-thermal contributions to the luminosity are negligible. To test this assumption and estimate the non-thermal contribution, we assume that the non-thermal flux can be estimated by combining the contribution from the luminosity of H$\alpha$, the radio luminosity, the UV luminosity, and the X-ray luminosity:
\begin{equation}
L_{\rm nonthermal} \approx L_{\rm H\alpha} + L_{\rm rad} + L_{\rm x},
\end{equation}
and adopt characteristic values for each of these based on detections from \citet{Hawley2003a} and \citet[][and references therein]{Berger2010} of  $L_{\rm H\alpha} = L_{\rm bol} \times  10^{-4}$, $L_{\rm rad} = L_{\rm bol}\times10^{-6}$, $L_{\rm UV} = L_{\rm C IV} \sim 0.1 L_{\rm H\alpha} \sim  L_{\rm bol} \times  10^{-5}$, and
$L_{\rm X} = L_{\rm bol} \times 10^{-4}$. This calculation results in an upper limit of
\begin{equation}
L_{\rm nonthermal} \approx L_{\rm bol} \times 10^{-4} + L_{\rm bol}\times10^{-6} + L_{\rm bol}\times10^{-5} + L_{\rm bol} \times 10^{-4} = 0.000211 \times L_{\rm bol},
\end{equation}
which is a 0.02\% non-thermal contribution to the bolometric luminosity. This result indicates that $L_{\rm bol} \approx L_{\rm thermal}$ is a reasonably accurate assumption, so we calculate the strength of H$\alpha$ using T$_{\rm eff}$ instead of explicitly estimating $R$ or $L_{bol}$:
\begin{equation}
\frac{L_{\rm H\alpha}}{L_{\rm bol}} = \frac{F_{\rm H\alpha} \times f }{\sigma T_{\rm eff}^4},
\end{equation}
Each photospheric model is generated for a specific T$_{\rm eff}$, and it is straightforward to explore the effects of varying a single parameter rather than both radius and luminosity. The uncertainties in the final calculation are characterized by varying T$_{\rm eff}$ by $\pm$200K; this dispersion is greater than the 0.02\% uncertainty from the non-thermal contribution to the bolometric luminosity and also characteristic of the uncertainty assigned to $T_{\rm eff}$ for a particular spectral type.

\LongTables
\begin{deluxetable*}{llllllll} \tablewidth{0pt}  \tabletypesize{\scriptsize}
\tablecaption{L Dwarf H$\alpha$ Detections and Emission Strength \label{tab:Lha} }
\tablehead{ \colhead{2MASS Designation}  &  \colhead{Spectral} & \colhead{H$\alpha$} & \colhead{EW H$\alpha$}  & \colhead{log(L$_{H\alpha}$/L$_{bol}$)} & \colhead{$\sigma_{\rm H\alpha}/\langle{\rm H\alpha}\rangle$} & \colhead{Fractional\tablenotemark{a}} & \colhead{Variable?} \\  \colhead{(2MASS J+)}  &  \colhead{Type} & \colhead{Ref.} & \colhead{(\AA)} & \colhead{} & \colhead{} & \colhead{Variability} & \colhead{}  
}
\startdata
00043484-4044058  &  L5  &  1  &  $<$0.20  &  $<$-6.60  &  1.1  &  6.5  &  y  \\
                  &  L5  &  2  &     1.50  &     -5.72  & & &   \\
00154476+3516026  &  L2  &  3  &     2.00  &     -5.45  &  \nodata  &  \nodata  &  \nodata  \\
00244419-2708242  &  L0  &  2  &  $<$0.38  &  $<$-6.08  &  \nodata  &  \nodata  &  \nodata  \\
00283943+1501418  &  L4.5  &  3  &  $<$2.00  &  $<$-5.57  &  \nodata  &  \nodata  &  \nodata  \\
00303013-1450333  &  L7  &  3  &  $<$10.0  &  $<$-5.04  &  \nodata  &  \nodata  &  \nodata  \\
00304384+3139321  &  L2  &  4  &     4.40  &     -5.10  &  \nodata  &  \nodata  &  \nodata  \\
00361617+1821104  &  L3.5  &  3  &  $<$0.50  &  $<$-6.12  &  \nodata  &  \nodata  &  n  \\
                  &  L3.5  &  1  &  $<$1.09  &  $<$-5.78  & & &   \\
00452143+1634446  &  L0  &  2  &     10.06  &     -4.66  &  \nodata  &  \nodata  &  \nodata  \\
00511078-1544169  &  L3.5  &  3  &  $<$2.00  &  $<$-5.51  &  \nodata  &  \nodata  &  \nodata  \\
00584253-0651239  &  L0  &  3  &     2.00  &     -5.36  &  \nodata  &  \nodata  &  \nodata  \\
01033203+1935361  &  L6  &  3  &  $<$1.00  &  $<$-5.96  &  \nodata  &  \nodata  &  \nodata  \\
01075242+0041563  &  L8  &  2  &  $<$15.43  &  $<$-4.94  &  \nodata  &  \nodata  &  \nodata  \\
01282664-5545343  &  L3  &  2  &  $<$6.82  &  $<$-4.96  &  \nodata  &  \nodata  &  \nodata  \\
01291221+3517580  &  L4  &  4  &  $<$0.50  &  $<$-6.14  &  \nodata  &  \nodata  &  \nodata  \\
01353586+1205216  &  L1.5  &  3  &     7.00  &     -4.88  &  \nodata  &  \nodata  &  \nodata  \\
01443536-0716142  &  L5  &  2  &  $<$3.06  &  $<$-5.41  &  1.0  &  7.2  &  y  \\
                  &  L5  &  5  &     25.00  &     -4.50  &  & &  \\
                  &  L5  &  5  &     6.00  &     -5.12  &  & &  \\
01473344+3453112  &  L0.5  &  4  &  $<$0.50  &  $<$-5.99  &  \nodata  &  \nodata  &  \nodata  \\
02050344+1251422  &  L5  &  3  &  $<$1.00  &  $<$-5.90  &  \nodata  &  \nodata  &  \nodata  \\
02052940-1159296  &  L7  &  4  &  $<$2.70  &  $<$-5.60  &  \nodata  &  \nodata  &  \nodata  \\
02081833+2542533  &  L1  &  3  &  $<$0.50  &  $<$-6.01  &  \nodata  &  \nodata  &  \nodata  \\
02082363+2737400  &  L5  &  3  &  $<$1.00  &  $<$-5.90  &  \nodata  &  \nodata  &  \nodata  \\
02085499+2500488  &  L5  &  3  &  $<$1.00  &  $<$-5.90  &  \nodata  &  \nodata  &  \nodata  \\
02132880+4444453  &  L1.5  &  2  &     2.35  &     -5.35  &  \nodata  &  \nodata  &  \nodata  \\
02243670+2537042  &  L2  &  3  &  $<$1.00  &  $<$-5.75  &  \nodata  &  \nodata  &  \nodata  \\
02284243+1639329  &  L0  &  2  &     3.49  &     -5.12  &  \nodata  &  \nodata  &  \nodata  \\
02355993-2331205  &  L1  &  1  &  $<$0.20  &  $<$-6.40  &  \nodata  &  \nodata  &  \nodata  \\
02424355+1607392  &  L1.5  &  4  &  $<$0.50  &  $<$-6.03  &  \nodata  &  \nodata  &  \nodata  \\
02511490-0352459  &  L3  &  2  &  $<$0.33  &  $<$-6.27  &  \nodata  &  \nodata  &  \nodata  \\
02550357-4700509  &  L8.0  &  1  &  $<$0.20  &  $<$-6.83  &  \nodata  &  \nodata  &  n  \\
                  &  L8  &  2  &  $<$1.00  &  $<$-6.13  &  & &  \\
02572581-3105523  &  L8  &  2  &  $<$2.00  &  $<$-5.83  &  \nodata  &  \nodata  &  \nodata  \\
03020122+1358142  &  L3  &  3  &  $<$2.00  &  $<$-5.49  &  \nodata  &  \nodata  &  \nodata  \\
03062684+1545137  &  L6  &  3  &  $<$6.00  &  $<$-5.18  &  \nodata  &  \nodata  &  \nodata  \\
03090888-1949387  &  L4.5  &  3  &  $<$7.00  &  $<$-5.02  &  \nodata  &  \nodata  &  \nodata  \\
03105986+1648155  &  L8  &  3  &  $<$3.00  &  $<$-5.65  &  \nodata  &  \nodata  &  \nodata  \\
03140344+1603056  &  L0  &  1  &     7.72  &     -4.77  &  1.2  &  9.0  &  y  \\
                  &  L0  &  2  &     0.77  &     -5.78  &  & &  \\
03261367+2950152  &  L3.5  &  4  &     9.10  &     -4.86  &  \nodata  &  \nodata  &  \nodata  \\
03284265+2302051  &  L8  &  3  &  $<$4.00  &  $<$-5.52  &  \nodata  &  \nodata  &  \nodata  \\
03370359-1758079  &  L4.5  &  3  &  $<$5.00  &  $<$-5.17  &  \nodata  &  \nodata  &  \nodata  \\
03440892+0111251  &  L1  &  6  &  $<$-1.09  &  $<$-5.67  &  \nodata  &  \nodata  &  \nodata  \\
03454316+2540233  &  L0  &  4  &  $<$0.30  &  $<$-6.18  &  \nodata  &  \nodata  &  \nodata  \\
03552337+1133437  &  L6  &  2  &  $<$24.29  &  $<$-4.57  &  \nodata  &  \nodata  &  \nodata  \\
03554191+2257016  &  L3  &  4  &  $<$0.50  &  $<$-6.09  &  \nodata  &  \nodata  &  \nodata  \\
04090950+2104393  &  L3  &  3  &  $<$1.00  &  $<$-5.79  &  \nodata  &  \nodata  &  \nodata  \\
04234858-0414035  &  L7  &  2  &  $<$10.45  &  $<$-5.02  &  \nodata  &  \nodata  &  \nodata  \\
04390101-2353083  &  L6.5  &  2  &  $<$6.32  &  $<$-5.20  &  \nodata  &  \nodata  &  \nodata  \\
04455387-3048204  &  L2  &  2  &  $<$0.75  &  $<$-5.87  &  \nodata  &  \nodata  &  \nodata  \\
05002100+0330501  &  L4  &  2  &  $<$8.46  &  $<$-4.91  &  \nodata  &  \nodata  &  \nodata  \\
05233822-1403022  &  L2.5  &  1  &     0.34  &     -6.24  &  \nodata  &  \nodata  &  n  \\
                  &  L2.5  &  2  &  $<$0.78  &  $<$-5.88  &  & &  \\
06023045+3910592  &  L1.0  &  1  &  $<$0.49  &  $<$-6.01  &  \nodata  &  \nodata  &  \nodata  \\
06244595-4521548  &  L5  &  2  &  $<$9.92  &  $<$-4.90  &  \nodata  &  \nodata  &  \nodata  \\
06523073+4710348  &  L4.5  &  2  &  $<$2.90  &  $<$-5.41  &  \nodata  &  \nodata  &  \nodata  \\
07003664+3157266  &  L3.5  &  1  &  $<$1.78  &  $<$-5.56  &  \nodata  &  \nodata  &  n  \\
                  &  L3.5  &  2  &  $<$1.16  &  $<$-5.75  &  & &  \\
07082133+2950350  &  L5  &  3  &  $<$2.00  &  $<$-5.60  &  \nodata  &  \nodata  &  \nodata  \\
07400966+3212032  &  L4.5  &  3  &  $<$2.00  &  $<$-5.57  &  \nodata  &  \nodata  &  \nodata  \\
07464256+2000321  &  L1  &  6  &     1.78  &     -5.45  &  0.1  &  0.3  &  n  \\
                  &  L0.5  &  1  &     2.36  &     -5.31  &  & &  \\
                  &  L0.5  &  3  &     2.00  &     -5.38  &  & &  \\
07533217+2917119  &  L2  &  3  &  $<$0.50  &  $<$-6.05  &  \nodata  &  \nodata  &  \nodata  \\
07562529+1244560  &  L6  &  3  &  $<$3.00  &  $<$-5.48  &  \nodata  &  \nodata  &  \nodata  \\
08014056+4628498  &  L6.5  &  3  &  $<$2.00  &  $<$-5.70  &  \nodata  &  \nodata  &  \nodata  \\
08111040+1855280  &  L1  &  6  &     4.38  &     -5.06  &  \nodata  &  \nodata  &  \nodata  \\
08202996+4500315  &  L5  &  3  &  $<$2.00  &  $<$-5.60  &  \nodata  &  \nodata  &  \nodata  \\
08251968+2115521  &  L7.5  &  1  &  $<$0.20  &  $<$-6.78  &  \nodata  &  \nodata  &  n  \\
                  &  L7.5  &  3  &  $<$2.00  &  $<$-5.78  &  & &  \\
08283419-1309198  &  L2.0  &  1  &  $<$0.20  &  $<$-6.45  &  \nodata  &  \nodata  &  \nodata  \\
08290664+1456225  &  L1  &  6  &     1.64  &     -5.49  &  0.8  &  2.3  &  y  \\
                  &  L2  &  3  &  $<$0.50  &  $<$-6.05  &  & &  \\
08295707+2655099  &  L6.5  &  3  &  $<$10.0  &  $<$-5.00  &  \nodata  &  \nodata  &  \nodata  \\
08300825+4828482  &  L8  &  2  &  $<$1.60  &  $<$-5.92  &  \nodata  &  \nodata  &  \nodata  \\
08320451-0128360  &  L1.5  &  3  &     2.00  &     -5.42  &  \nodata  &  \nodata  &  \nodata  \\
08354256-0819237  &  L5.0  &  1  &  $<$0.20  &  $<$-6.60  &  \nodata  &  \nodata  &  n  \\
                  &  L5  &  2  &  $<$3.54  &  $<$-5.35  &  & &  \\
08355829+0548308  &  L3  &  6  &     5.07  &     -5.09  &  \nodata  &  \nodata  &  \nodata  \\
08472872-1532372  &  L2  &  2  &  $<$1.11  &  $<$-5.70  &  \nodata  &  \nodata  &  \nodata  \\
08503593+1057156  &  L6  &  4  &  $<$0.90  &  $<$-6.01  &  \nodata  &  \nodata  &  \nodata  \\
09083803+5032088  &  L7  &  2  &  $<$5.60  &  $<$-5.29  &  \nodata  &  \nodata  &  \nodata  \\
09111297+7401081  &  L0  &  2  &  $<$3.31  &  $<$-5.14  &  \nodata  &  \nodata  &  \nodata  \\
09130320+1841501  &  L3  &  4  &     0.80  &     -5.89  &  0.8  &  3.0  &  y  \\
                  &  L3.0  &  1  &  $<$0.20  &  $<$-6.49  &  \nodata  &  \nodata  &  \nodata  \\
09153413+0422045  &  L7  &  2  &  $<$5.29  &  $<$-5.31  &  \nodata  &  \nodata  &  \nodata  \\
09183815+2134058  &  L2.5  &  4  &  $<$0.30  &  $<$-6.29  &  \nodata  &  \nodata  &  \nodata  \\
09201223+3517429  &  L6.5  &  3  &  $<$0.50  &  $<$-6.30  &  \nodata  &  \nodata  &  \nodata  \\
09211410-2104446  &  L2  &  2  &  $<$2.99  &  $<$-5.27  &  \nodata  &  \nodata  &  n  \\
                  &  L2.0  &  1  &  $<$0.32  &  $<$-6.24  &  & &  \\
09283972-1603128  &  L2  &  3  &  $<$0.50  &  $<$-6.05  &  \nodata  &  \nodata  &  \nodata  \\
09293364+3429527  &  L8  &  3  &  $<$3.00  &  $<$-5.65  &  \nodata  &  \nodata  &  \nodata  \\
09440279+3131328  &  L2  &  3  &  $<$1.00  &  $<$-5.75  &  \nodata  &  \nodata  &  \nodata  \\
09510549+3558021  &  L6  &  3  &  $<$5.00  &  $<$-5.26  &  \nodata  &  \nodata  &  \nodata  \\
10170754+1308398  &  L2  &  6  &     6.79  &     -4.91  &  \nodata  &  \nodata  &  \nodata  \\
10224821+5825453  &  L1  &  2  &     24.00  &     -4.32  &  1.1  &  6.2  &  y  \\
                  &  L1  &  1  &     3.32  &     -5.18  &  & &  \\
10292165+1626526  &  L2.5  &  3  &     0.50  &     -6.07  &  0.8  &  2.9  &  y  \\
                  &  L2.5  &  1  &     1.96  &     -5.48  &  & &  \\
10352455+2507450  &  L1  &  3  &  $<$1.00  &  $<$-5.70  &  \nodata  &  \nodata  &  \nodata  \\
10430758+2225236  &  L8  &  2  &  $<$4.60  &  $<$-5.46  &  \nodata  &  \nodata  &  \nodata  \\
10452400-0149576  &  L1.0  &  1  &  $<$0.20  &  $<$-6.40  &  \nodata  &  \nodata  &  n  \\
                  &  L1  &  2  &  $<$17.68  &  $<$-4.46  &  & &  \\
10473109-1815574  &  L2.5  &  1  &  $<$1.16  &  $<$-5.70  &  \nodata  &  \nodata  &  \nodata  \\
10484281+0111580  &  L1  &  6  &     4.28  &     -5.07  &  0.9  &  3.0  &  y  \\
                  &  L1  &  2  &  $<$1.11  &  $<$-5.66  &  & &  \\
                  &  L1  &  1  &     1.08  &     -5.67  &  & &  \\
10511900+5613086  &  L1  &  6  &     1.54  &     -5.52  &  \nodata  &  \nodata  &  n  \\
                  &  L2  &  2  &  $<$2.31  &  $<$-5.38  &  & &  \\
10515129+1311164  &  L0  &  6  &     1.39  &     -5.52  &  \nodata  &  \nodata  &  \nodata  \\
10584787-1548172  &  L3  &  4  &     1.60  &     -5.59  &  \nodata  &  \nodata  &  \nodata  \\
11023375-2359464  &  L4.5  &  3  &  $<$5.00  &  $<$-5.17  &  \nodata  &  \nodata  &  \nodata  \\
11040127+1959217  &  L4  &  2  &  $<$1.62  &  $<$-5.63  &  \nodata  &  \nodata  &  \nodata  \\
11083081+6830169  &  L0.5  &  2  &  $<$0.86  &  $<$-5.75  &  \nodata  &  \nodata  &  \nodata  \\
11122567+3548131  &  L4.5  &  3  &  $<$1.00  &  $<$-5.87  &  \nodata  &  \nodata  &  \nodata  \\
11235564+4122286  &  L2.5  &  3  &  $<$1.00  &  $<$-5.77  &  \nodata  &  \nodata  &  \nodata  \\
11455714+2317297  &  L1.5  &  4  &     4.20  &     -5.10  &  0.1  &  0.1  &  n  \\
                  &  L1.5  &  1  &     3.69  &     -5.16  &  & &  \\
11463449+2230527  &  L3  &  6  &     1.68  &     -5.56  &  1.0  &  4.6  &  y  \\
                  &  L3  &  4  &  $<$0.30  &  $<$-6.31  &  & &  \\
11550087+2307058  &  L4  &  4  &  $<$1.00  &  $<$-5.84  &  \nodata  &  \nodata  &  \nodata  \\
11553952-3727350  &  L2  &  2  &  $<$2.91  &  $<$-5.28  &  \nodata  &  \nodata  &  n  \\
                  &  L2  &  1  &     1.00  &     -5.75  &  & &  \\
11593850+0057268  &  L0  &  6  &     1.66  &     -5.44  &  0.5  &  1.0  &  n  \\
                  &  L0  &  1  &     3.31  &     -5.14  &  & &  \\
12025009+4204531  &  L0  &  6  &     1.86  &     -5.39  &  \nodata  &  \nodata  &  \nodata  \\
12035812+0015500  &  L3  &  1  &  $<$1.17  &  $<$-5.72  &  \nodata  &  \nodata  &  n  \\
                  &  L3  &  2  &  $<$10.85  &  $<$-4.76  &  & &  \\
12043036+3212595  &  L0  &  6  &     1.02  &     -5.65  &  \nodata  &  \nodata  &  \nodata  \\
12130336-0432437  &  L5  &  2  &  $<$2.52  &  $<$-5.49  &  \nodata  &  \nodata  &  \nodata  \\
12212770+0257198  &  L0  &  6  &     6.05  &     -4.88  &  0.1  &  0.2  &  n  \\
                  &  L0  &  2  &     5.96  &     -4.89  &  & &  \\
                  &  L0  &  1  &     5.01  &     -4.96  &  & &  \\
12281523-1547342  &  L5  &  4  &  $<$0.60  &  $<$-6.12  &  \nodata  &  \nodata  &  \nodata  \\
12392727+5515371  &  L5  &  3  &  $<$3.00  &  $<$-5.42  &  \nodata  &  \nodata  &  \nodata  \\
12464678+4027150  &  L4  &  3  &  $<$1.00  &  $<$-5.84  &  \nodata  &  \nodata  &  \nodata  \\
13004255+1912354  &  L1  &  2  &  $<$0.40  &  $<$-6.10  &  \nodata  &  \nodata  &  n  \\
                  &  L1  &  1  &  $<$0.20  &  $<$-6.40  &  & &  \\
13054019-2541059  &  L2  &  4  &     1.90  &     -5.47  &  0.1  &  0.1  &  n  \\
                  &  L2  &  1  &     1.75  &     -5.50  &  & &  \\
13153094-2649513  &  L5  &  7  &     121.00  &     -3.81  &  0.7  &  3.8  &  y  \\
                  &  L5  &  8  &     57.00  &     -4.14  &  & &  \\
                  &  L5  &  7  &     25.00  &     -4.50  &  & &  \\
13285503+2114486  &  L5  &  4  &  $<$2.00  &  $<$-5.60  &  \nodata  &  \nodata  &  \nodata  \\
13313310+3407583  &  L0  &  6  &     1.03  &     -5.65  &  \nodata  &  \nodata  &  \nodata  \\
13322863+2635079  &  L2  &  3  &  $<$2.00  &  $<$-5.45  &  \nodata  &  \nodata  &  \nodata  \\
13340623+1940351  &  L1.5  &  1  &  $<$0.20  &  $<$-6.42  &  1.3  &  20.0  &  y  \\
                  &  L1.5  &  4  &     4.20  &     -5.10  &  \nodata  &  \nodata  &  \nodata  \\
13382615+4140342  &  L2.5  &  3  &  $<$1.00  &  $<$-5.77  &  \nodata  &  \nodata  &  \nodata  \\
13384944+0437315  &  L1  &  6  &     7.80  &     -4.81  &  \nodata  &  \nodata  &  \nodata  \\
13422362+1751558  &  L2.5  &  4  &  $<$2.20  &  $<$-5.43  &  \nodata  &  \nodata  &  \nodata  \\
13431670+3945087  &  L5  &  3  &  $<$3.00  &  $<$-5.42  &  \nodata  &  \nodata  &  \nodata  \\
13595510-4034582  &  L1  &  1  &  $<$0.20  &  $<$-6.40  &  \nodata  &  \nodata  &  \nodata  \\
14111735+3936363  &  L1.5  &  3  &  $<$1.00  &  $<$-5.73  &  \nodata  &  \nodata  &  \nodata  \\
14122449+1633115  &  L0.5  &  3  &     4.00  &     -5.08  &  0.7  &  1.8  &  y  \\
                  &  L0.5  &  1  &     1.45  &     -5.52  &  & &  \\
14162408+1348263  &  L6  &  6  &  $<$-0.11  &  $<$-6.91  &  \nodata  &  \nodata  &  \nodata  \\
14213145+1827407  &  L0  &  2  &     2.04  &     -5.35  &  \nodata  &  \nodata  &  \nodata  \\
14243909+0917104  &  L4  &  4  &  $<$0.80  &  $<$-5.94  &  \nodata  &  \nodata  &  \nodata  \\
14252798-3650229  &  L3  &  2  &  $<$5.71  &  $<$-5.03  &  \nodata  &  \nodata  &  \nodata  \\
14284323+3310391  &  L0  &  6  &     3.49  &     -5.12  &  \nodata  &  \nodata  &  \nodata  \\
14385498-1309103  &  L3  &  3  &  $<$4.00  &  $<$-5.19  &  \nodata  &  \nodata  &  \nodata  \\
14392836+1929149  &  L1  &  1  &     3.48  &     -5.16  &  0.9  &  10.6  &  y  \\
                  &  L1  &  4  &  $<$0.30  &  $<$-6.23  &  & &  \\
                  &  L1  &  2  &  $<$1.36  &  $<$-5.57  &  & &  \\
14394092+1826369  &  L1  &  4  &  $<$0.60  &  $<$-5.93  &  \nodata  &  \nodata  &  \nodata  \\
14413716-0945590  &  L0.5  &  1  &  $<$1.53  &  $<$-5.50  &  \nodata  &  \nodata  &  \nodata  \\
14482563+1031590  &  L5  &  2  &  $<$8.33  &  $<$-4.98  &  \nodata  &  \nodata  &  \nodata  \\
15065441+1321060  &  L3  &  1  &     0.69  &     -5.95  &  \nodata  &  \nodata  &  n  \\
                  &  L3  &  2  &  $<$7.87  &  $<$-4.89  &  & &  \\
15074769-1627386  &  L5  &  3  &  $<$0.50  &  $<$-6.20  &  \nodata  &  \nodata  &  n  \\
                  &  L5  &  1  &  $<$0.29  &  $<$-6.43  &  & &  \\
15150083+4847416  &  L6  &  2  &  $<$10.11  &  $<$-4.96  &  \nodata  &  \nodata  &  \nodata  \\
15232263+3014562  &  L8  &  3  &  $<$15.00  &  $<$-4.95  &  \nodata  &  \nodata  &  \nodata  \\
15394189-0520428  &  L3.5  &  2  &  $<$2.77  &  $<$-5.37  &  \nodata  &  \nodata  &  \nodata  \\
15525906+2948485  &  L0  &  6  &     1.50  &     -5.49  &  \nodata  &  \nodata  &  \nodata  \\
15532142+2109071  &  L5.5  &  4  &  $<$4.30  &  $<$-5.29  &  \nodata  &  \nodata  &  \nodata  \\
15551573-0956055  &  L1  &  1  &     2.45  &     -5.32  &  \nodata  &  \nodata  &  \nodata  \\
15564434+1723089  &  L1  &  6  &     1.19  &     -5.63  &  \nodata  &  \nodata  &  \nodata  \\
16000548+1708328  &  L1.5  &  3  &  $<$1.50  &  $<$-5.55  &  \nodata  &  \nodata  &  \nodata  \\
16134557+1708273  &  L0  &  6  &     4.37  &     -5.02  &  \nodata  &  \nodata  &  \nodata  \\
16154416+3559005  &  L3  &  1  &  $<$1.50  &  $<$-5.61  &  \nodata  &  \nodata  &  n  \\
                  &  L3  &  3  &  $<$1.00  &  $<$-5.79  &  & &  \\
16322911+1904407  &  L8  &  4  &  $<$4.00  &  $<$-5.52  &  \nodata  &  \nodata  &  \nodata  \\
16325610+3505076  &  L1  &  6  &  $<$0.31  &  $<$-6.22  &  \nodata  &  \nodata  &  \nodata  \\
16452211-1319516  &  L1.5  &  1  &     1.51  &     -5.55  &  \nodata  &  \nodata  &  \nodata  \\
16561885+2835056  &  L4.5  &  3  &  $<$6.00  &  $<$-5.09  &  \nodata  &  \nodata  &  \nodata  \\
16580380+7027015  &  L1  &  2  &  $<$13.55  &  $<$-4.57  &  \nodata  &  \nodata  &  \nodata  \\
17054834-0516462  &  L4  &  1  &  $<$0.20  &  $<$-6.54  &  \nodata  &  \nodata  &  \nodata  \\
17072343-0558249  &  L0  &  2  &  $<$3.43  &  $<$-5.13  &  \nodata  &  \nodata  &  \nodata  \\
17111353+2326333  &  L0  &  6  &     4.90  &     -4.97  &  \nodata  &  \nodata  &  \nodata  \\
17114573+2232044  &  L6.5  &  3  &  $<$4.00  &  $<$-5.39  &  \nodata  &  \nodata  &  \nodata  \\
17210390+3344160  &  L3  &  2  &  $<$1.89  &  $<$-5.51  &  \nodata  &  \nodata  &  \nodata  \\
17260007+1538190  &  L2  &  3  &  $<$2.00  &  $<$-5.45  &  \nodata  &  \nodata  &  \nodata  \\
17281150+3948593  &  L7  &  3  &  $<$7.00  &  $<$-5.19  &  \nodata  &  \nodata  &  \nodata  \\
17312974+2721233  &  L0  &  2  &     5.98  &     -4.88  &  0.0  &  0.0  &  n  \\
                  &  L0  &  1  &     5.99  &     -4.88  &  & &  \\
17434148+2127069  &  L2.5  &  3  &     6.00  &     -4.99  &  \nodata  &  \nodata  &  \nodata  \\
17534518-6559559  &  L4  &  2  &  $<$26.65  &  $<$-4.41  &  \nodata  &  \nodata  &  \nodata  \\
18071593+5015316  &  L1.5  &  1  &     3.79  &     -5.15  &  0.7  &  1.7  &  y  \\
                  &  L1.5  &  2  &     1.38  &     -5.59  &  & &  \\
18410861+3117279  &  L4  &  3  &  $<$2.00  &  $<$-5.54  &  \nodata  &  \nodata  &  \nodata  \\
18544597+8429470  &  L0.0  &  1  &     7.06  &     -4.81  &  \nodata  &  \nodata  &  \nodata  \\
19360187-5502322  &  L4  &  2  &  $<$7.77  &  $<$-4.95  &  \nodata  &  \nodata  &  \nodata  \\
20282035+0052265  &  L3  &  2  &  $<$6.33  &  $<$-4.99  &  \nodata  &  \nodata  &  \nodata  \\
20360316+1051295  &  L3  &  2  &  $<$6.34  &  $<$-4.99  &  \nodata  &  \nodata  &  \nodata  \\
20543585+1519043  &  L1  &  3  &  $<$6.00  &  $<$-4.93  &  \nodata  &  \nodata  &  \nodata  \\
20571538+1715154  &  L1.5  &  3  &  $<$2.00  &  $<$-5.42  &  \nodata  &  \nodata  &  \nodata  \\
20575409-0252302  &  L1.5  &  2  &     8.44  &     -4.80  &  0.0  &  0.0  &  n  \\
                  &  L1.5  &  1  &     8.15  &     -4.81  &  & &  \\
21011544+1756586  &  L7.5  &  3  &  $<$7.00  &  $<$-5.23  &  \nodata  &  \nodata  &  \nodata  \\
21041491-1037369  &  L3  &  1  &  $<$1.53  &  $<$-5.61  &  \nodata  &  \nodata  &  n  \\
                  &  L2.5  &  2  &  $<$3.41  &  $<$-5.24  &  & &  \\
22000201-3038327  &  L0  &  1  &     3.56  &     -5.11  &  \nodata  &  \nodata  &  \nodata  \\
22064498-4217208  &  L2  &  3  &  $<$2.00  &  $<$-5.45  &  \nodata  &  \nodata  &  \nodata  \\
22081363+2921215  &  L2  &  3  &  $<$2.00  &  $<$-5.45  &  \nodata  &  \nodata  &  \nodata  \\
22244381-0158521  &  L4.5  &  3  &     1.00  &     -5.87  &  \nodata  &  \nodata  &  n  \\
                  &  L4.5  &  1  &  $<$1.21  &  $<$-5.78  &  & &  \\
22521073-1730134  &  L5.5  &  2  &  $<$173.70  &  $<$-3.69  &  \nodata  &  \nodata  &  \nodata  \\
22551861-5713056  &  L3  &  2  &  $<$20.33  &  $<$-4.48  &  \nodata  &  \nodata  &  \nodata  \\
23254530+4251488  &  L8  &  2  &  $<$1.39  &  $<$-5.98  &  \nodata  &  \nodata  &  \nodata  \\
23352640+0817214  &  L0  &  6  &     3.82  &     -5.08  &  \nodata  &  \nodata  &  \nodata  \\
23515044-2537367  &  L0.5  &  1  &     2.76  &     -5.24  &  \nodata  &  \nodata  &  \nodata  
\enddata
\tablerefs{(1) \citet{Reiners2008}; (2) \citet{Schmidt2007}; (3) \citet{Kirkpatrick2000}; (4) \citet{Kirkpatrick1999}; (5) \citet{Liebert2003}; (6) this paper; (7) \citet{Hall2002}; (8) \citet{Burgasser2011b}}
\tablenotetext{a}{Fractional variability is defined as the total range of EW H$\alpha$ divided by the minimum EW H$\alpha$.}
\end{deluxetable*}

\end{document}